\documentclass[a4 paper, 11pt]{article}
\usepackage{jheppub} % for details on the use of the package, please see the JINST-author-manual
\usepackage{lineno}
\usepackage{braket}
\usepackage{slashed}
\usepackage{bbold}
\usepackage{amsmath}
\allowdisplaybreaks
\newcommand{\dd}{\mathrm{d}}
\newcommand{\be}{\begin{equation}}
\newcommand{\ee}{\end{equation}}

\newcommand{\nn}{\text}
\title{\boldmath New non-supersymmetric flux vacua from generalised calibrations}

\author{Vincent Menet}
\affiliation{Laboratoire de Physique Théorique et Hautes Energies - LPTHE\\
Sorbonne Université, 4 Place Jussieu, 75005 Paris, France}

\emailAdd{vmenet@lpthe.jussieu.fr}

\abstract{We construct a new class of non-supersymmetric ten-dimensional type II flux vacua, by studying first order differential equations which are deformations of the $\mathcal{N}=1$ supersymmetry conditions. We do so within the context of Generalised Complex Geometry, where there is a natural interpretation of the $\mathcal{N}=1$ supersymmetry conditions in terms of calibration conditions for probe D-branes, called D-string, domain-wall or space-filling branes, depending on them wrapping two, three or four non-compact dimensions. We focus on the class of non-supersymmetric vacua violating the D-string calibration condition, and write down their general equations of motion in the language of pure spinors. We solve them for a subclass of vacua, where the deformation of the calibration condition is dictated by the foliated geometry of the internal space. We also construct backgrounds violating both the domain-wall and D-string calibration conditions, generalising the one-parameter DWSB class of backgrounds introduced in L\"ust et al. We present several explicit solutions with SU$(2)$ and SU$(3)$ structures, and we investigate briefly their associated low energy effective theories.}

\begin{document}
\maketitle
\flushbottom
\newpage \section{Introduction}     

The exploration of the landscape of four-dimensional string compactifications has been mostly focused on vacua preserving at least $\mathcal{N}=1$ supersymmetry. One  reason is practical:  solving the supersymmetry conditions, which are first order differential equations, plus the Bianchi identities for the fluxes, guarantees to have solutions  to the full set of string or supergravity equations of motion. Without this way out, handling the equations of motion upfront is very hard, even in the supergravity approximation, since they are cumbersome second order differential equations.

There are also physical considerations motivating the study of supersymmetric string compactifications, namely the expectation that supersymmetry should be broken at energies smaller than the compactification scale. 

\medskip

Even if low energy supersymmetry breaking is a 
phenomenologically motivated scenario, in principle  nothing prevents supersymmetry from being spontaneously broken at arbitrarily high energies. In this paper, we consider this possibility, and focus on this much less studied corner of the string compactification landscape, worth exploring per se.

More precisely, we construct new classes of non-supersymmetric type II supergravity solutions  by breaking supersymmetry in a controlled way. We deform the conditions for $\mathcal{N}=1$ supersymmetry by adding supersymmetry breaking terms, which are controlled by small parameters, whose vanishing would restore supersymmetry. 

\medskip

The motivation behind this approach is to preserve some of the convenient features of supersymmetric vacua, mainly the possibility to use first order differential equations. Since supersymmetry is broken, in order to find solutions we have to make sure that the equations of motion  are satisfied. 
The goal is to find specific deformations of the BPS equations such that the additional constraints to impose in order to solve the equations of motion are manageable.

\medskip

We will use the framework of 
Generalised Complex Geometry, where the $\mathcal{N}=1$ BPS conditions have an interpretation in terms of calibration conditions of different probe D-branes. The $\mathcal{N}=1$ supersymmetry conditions for warped compactifications can be recast in a set of three differential equations on polyforms defined only on the internal compactification space \cite{Grana:2005sn}. Each of these three conditions can be interpreted as the conditions for calibrated D-brane probes in the geometry \cite{Martucci:2005ht}:  branes filling all the external space and branes that are domain-wall or string-like. 

In this language,  one can identify different supersymmetry breaking terms depending on which calibration condition is modified. In this paper 
we will always assume that space-filling branes are  calibrated\footnote{Let us stress here that the interpretation of the BPS conditions in terms of D-brane calibrations doesn't mean that the backgrounds have to have D-string, domain-wall or space-filling D-branes, but we impose the presence of space-filling D-branes for model building considerations.}, while we will allow the calibrations of D-strings and domain-wall branes to be violated. 

 \medskip
 
 A famous example of non-supersymmetric type IIB solutions that violate the domain-wall calibration condition are the GKP solutions \cite{Giddings:2001yu}, describing flux compactifications to four-dimensional Minkowski space with D3 and O3 sources, where supersymmetry is broken by the $H_{(0,3)}$ components of the NSNS-flux. The GKP backgrounds have been described within Generalised Complex Geometry in \cite{Lust:2008zd} as specific examples of a  general framework to describe non-supersymmetric solutions. 
 
The Generalised Complex Geometry description of the GKP backgrounds also offers an insightful geometrical interpretation of the domain-wall supersymmetry breaking term: it is given by 
 the current associated to the D3-branes in the background\footnote{Strictly speaking it is the smeared version of the generalised current associated to the D3-branes, as we will discus in the text.}.

\medskip

In the literature there is another example of non-supersymmetric solution, this time in type IIA,  \cite{Legramandi:2019ulq}, which in the language of Generalised Complex Geometry corresponds to the violation of  the D-string calibration condition,  where supersymmetry is again violated through additional NSNS-flux components with respect to the supersymmetric case, but there is no further geometrical interpretation of the corresponding supersymmetry breaking term. Moreover, the question of stability of such backgrounds remains unaddressed. 

In this paper, we want to extend the study of non-supersymmetric vacua violating  the D-string calibration condition. More precisely, we will construct new non-supersymmetric type II solutions, where the current associated to the space-filling D-branes present in our backgrounds will serve as a building block for the supersymmetry breaking term violating the D-string calibration condition, in a sense that will be made precise in the text.

The motivation behind this construction is two-fold. The first one is simplicity: defining  supersymmetry breaking in terms of the current of the background's D-branes is a natural and simple ansatz, which in turn reduces the equations of motion to a reasonable set of additional constraints. The second reason is that it can be useful to address the question of stability of these non-supersymmetric vacua: 
in Generalised Complex Geometry, D-branes current can enter the effective potential associated to a given ten-dimensional background, and are particularly useful as they allow to use powerful positivity arguments from the branes calibration bounds in the study of the effective potential. 

We can show that our new class of vacua shares one interesting property with the GKP backgrounds, namely the fact that there is a natural truncation of the ten-dimensional theory, suggested by the geometry, such that the off-shell effective potential is positive semi-definite, and vanishes at the solutions. This statement is however not quite equivalent to claiming the stability of these new vacua, since we have limited control on the aforementioned truncation, as we will discuss at length.

We will also construct a new class of backgrounds generalising the GKP vacua, where both  the domain-wall and D-string calibration conditions are violated.

\medskip

The outline of the paper is as follows. In Section \ref{sec:gcg}, we briefly recall how the $\mathcal{N}=1$ supersymmetry conditions for warped compactifications are recast in the language of  Generalised Complex Geometry
and how ordinary calibrations can be extended to this framework to describe supersymmetric  D-brane probes in these backgrounds.  In Section \ref{sec:n0gcg} we review the Generalised Complex Geometry description of the GKP-like backgrounds and we introduce our two new classes of non-supersymmetric backgrounds, with pure D-string supersymmetry breaking and mixed Domain-wall and D-string supersymmetry breaking. In Section \ref{sec:effpot}, we write the effective potential for these compactifications and derive the equations of motion in the Generalised Complex Geometry formalism, first for completely general D-string supersymmetry breaking, and then for the two new classes of  backgrounds we found. We also address the question of the stability of these new solutions. Finally, in Section \ref{sec:ex}, which can be read (almost) independently, we present different explicit examples of new non-supersymmetric vacua with SU$(2)$ and SU$(3)$ structures.

\section{$\mathcal{N}=1$ flux vacua in Generalised Complex Geometry}
\label{sec:gcg}

\subsection{Compactification and pure spinors}
   We consider type II supergravity backgrounds that are the warped product of four-dimensional Minkowski space $X_4$ and a six-dimensional compact manifold $\mathcal{M}$, with the following metric ansatz
\be 
\label{10dmet}
\dd s^2_{10}=e^{2A(y)}\eta_{\mu\nu}\dd x^\mu\dd x^\nu+g_{mn}\dd y^m\dd y^n,
\ee
where $x^\mu$, $\mu=0,...,3$ are the external coordinates on $X_4$, and $y^m$, $m=1,...,6$ are the coordinates on $\mathcal{M}$. 

The Poincar\'e invariance of $X_4$  constrains the  NSNS and RR-fluxes: the NSNS-field-strength $H$ can only have internal legs, and the ten-dimensional RR-field-strength must take the form\footnote{We use the  democratic formulation of \cite{Bergshoeff:2001pv}, where 
$$F^{10}=\sum_q F^{10}_q $$
with $q=0,2,...10 $ for type IIA and $q=1,3,...9$ for type IIB. }
\be F^{10}=F+e^{4A}\text{vol}_4\wedge\Tilde{F},\label{RR}\ee
where $F$ and $\Tilde{F}$ are purely internal and are related by the self duality of $F^{10}$ as
    \begin{equation}
         \Tilde{F}=\Tilde{\ast}_6 F = \ast_6 \sigma (F)\label{fselfdual}
    \end{equation} 
with $\sigma$ the reversal of all form indices. 

\medskip

In this section  we focus on backgrounds preserving $\mathcal{N}=1$ supersymmetry,  which amounts to the vanishing of the following gravitino and dilatino variations
  \begin{align}
  \label{10dsusyg}
\delta_\epsilon\psi_M=\left(\nabla_M+\frac{1}{4}\iota_M\slashed{H}\sigma_3+\frac{e^\phi}{16} \begin{pmatrix}
    0 & \slashed{F}^{10}\\
    -\sigma(\slashed{F}^{10}) & 0
\end{pmatrix}\Gamma_M\Gamma_{(10)}\right) \begin{pmatrix}
           \epsilon_1 \\
          \epsilon_{2}
         \end{pmatrix}\\
  \label{10dsusyd}
\delta_\epsilon\lambda=\left(\slashed{\partial}\phi+\frac{1}{2}\slashed{H}\sigma_3+\frac{e^\phi}{16} \Gamma^M\begin{pmatrix}
    0 & \slashed{F}^{10}\\
    -\sigma(\slashed{F}^{10}) & 0
\end{pmatrix}\Gamma_M\Gamma_{(10)}\right)\begin{pmatrix}
           \epsilon_1 \\
          \epsilon_{2}
         \end{pmatrix},
  \end{align}
where $\epsilon_1$ and $\epsilon_2$ are ten-dimensional Majorana-Weyl  spinors, and for a $p$-form $\omega$ the slash symbol denotes
\begin{equation}
\slashed{\omega}=\frac{1}{p!}\omega_{M_1...M_p}\Gamma^{M_1...M_p}. 
\end{equation}

\medskip
    
For backgrounds of warped type the spinors $\epsilon_1$ and $\epsilon_2$ split in the following way
  \begin{equation}
\epsilon_1=\zeta\otimes\eta_{1}+c.c.\qquad \epsilon_2=\zeta\otimes\eta_{2} +c.c.\label{10dspinors}
\end{equation}
where $\zeta$ is a Weyl spinor of positive chirality on $X_4$, and 
$\eta_{1}$ and $\eta_{2}$ are Weyl spinors on the six-dimensional internal space. $\eta_{1}$ has positive chirality, while  $\eta_2$
has negative chirality in type IIA and positive chirality in type IIB.  
In the rest of the paper we will assume that the spinors $\eta_1$ and $\eta_2$ are globally defined such that each define an SU$(3)$ structure on $\mathcal{M}$.

\medskip

The vanishing of the supersymmetry variations  \eqref{10dsusyg}, \eqref{10dsusyd} is equivalent to the set of differential equations \cite{Grana:2005sn}
\begin{align}
  \dd_H(\text{e}^{3A-\phi}\Psi_2)=& 0 \label{ret}\\ 
  \dd_H(\text{e}^{2A-\phi}\text{Im}\Psi_1)=&\ 0\label{second}\\
   \dd_H(e^{4A-\phi}\text{Re}\Psi_1)=&\ e^{4A}\Tilde{\ast}_6 F\label{susy3}
\end{align}
where $\dd_H=\dd+H\wedge$, and  $\Psi_1$ and $\Psi_2$ are polyforms 
defined from the internal supersymmetry parameters\footnote{Strictly speaking one should think of these tensor product in terms of the following Fierz identity \be \eta\otimes\chi=\sum_{k=0}^6 \frac{1}{k!}\left(\chi^\dagger\gamma_{m_k...m_1}\eta\right)\gamma^{m_1...m_k}.\ee These tensor products are then isomorphic to polyforms through the Clifford map \eqref{cliffmap}, so we treat them as such from now on. The conventions for the internal gamma matrices are given in Appendix \ref{sec:AppA}.}\begin{align}
        \Psi_1=&-\frac{8i}{\lVert \eta \rVert^2}\eta_1\otimes\eta_2^\dagger\\
         \Psi_2=&-\frac{8i}{\lVert \eta \rVert^2}\eta_1\otimes\eta_2^T.
    \end{align}
The polyforms $\Psi_1$ and $\Psi_2$ are odd/even and even/odd in type IIA/IIB, respectively
\be \Psi_1=\Psi_\mp,\quad \Psi_2=\Psi_\pm \, . \ee

The previous equations have a nice interpretation  within the framework of Generalised Complex Geometry (GCG). 
In Generalised Complex Geometry, one replaces the tangent bundle of the internal manifold with a generalised tangent bundle  
\begin{equation}
    E\simeq T\mathcal{M}\oplus T^\ast\mathcal{M} \, , 
\end{equation} 
whose sections are locally sums of vectors and one-forms (see Appendix \ref{sec:AppB} for more details)
\begin{equation}
   V=v+\xi\in\Gamma(E).
\end{equation}
In this language, the polyforms $\Psi_1$ and $\Psi_2$ are interpreted as ${\rm Cliff(6,6)}$ pure spinors, which, by construction, are globally defined and compatible
\begin{align}
&\braket{\Psi_1,V\cdot\Psi_2}=\braket{\Psi_1,V\cdot\bar{\Psi}_2}=0\quad\forall \ V\in E\\
&\braket{\Psi_1,\bar{\Psi}_1}=\braket{\Psi_2,\bar{\Psi}_2}=-8i\text{vol}_6,
    \end{align}
where $\braket{\ ,\ }$ denotes the Mukai pairing 
\begin{equation}
\braket{\omega,\chi}=\omega\wedge\sigma(\chi)|_6,\label{Mukai}
\end{equation}
and $\cdot$ is the Clifford action of the generalised vector $V$  on generalised spinors defined in \eqref{vecspinoract}.

The natural inner product between two generalised vectors $V=v+\xi\in\Gamma(E)$ and $W=  w+\rho\in\Gamma(E)$  
\be 
   (V, W):=\iota_v\rho+\iota_w\xi
   \ee
defines an SO$(6,6)$ structure on $E$, which is reduced to SU$(3)\times$SU$(3)$ by the existence of the globally defined compatible spinors $\Psi_1$ and $\Psi_2$ \cite{Grana:2005sn}.

As discussed in more details in \cite{Gualtieri:2007ng}, a pure spinor $\Psi$ is associated to an almost generalised complex structure, $\mathcal{J}$, whose integrability is equivalent to the following differential condition on the pure spinor 
\begin{equation}
\dd \Psi  =  V\cdot \Psi \, , 
\end{equation}
for some $V\in T\mathcal{M}\oplus T^\ast\mathcal{M}$. 

Thus the supersymmetry condition \eqref{ret} can be interpreted as the integrability of the generalised almost complex structure associated to the pure spinor $\Psi_2$, while the integrability of the structure associated to $\Psi_1$ is obstructed by the RR-fluxes. 

This interpretation also allows a generalisation of the Hodge decomposition of tensors into holomorphic and anti-holomorphic components. For instance, the generalised spinor bundle decomposes as 
\be 
\Lambda^\bullet T^\ast\mathcal{M}\otimes\mathbb{C}=\bigoplus_{k=-3}^{3} V_k,
\ee
where $V_k$ are subbundles of fixed eigenvalue of the generalised complex structure $\mathcal{J}_2$ associated to $\Psi_2$. Both $\mathcal{J}_2$ and $V_k$ are defined in Appendix \ref{sec:AppB}. 
Moreover, the integrability of the 
 almost generalised complex structure $\mathcal{J}_2$ results in the following decomposition of  the exterior derivative
\be 
\dd_H\ :\ \Gamma(V_k)\rightarrow \Gamma(V_{k-1})\oplus\Gamma(V_{k+1}).\label{extdersusy}
\ee

\subsection{Generalised calibrations and supersymmetry}\label{subsec:calib}

The supersymmetry equations \eqref{ret}, \eqref{second} and \eqref{susy3} also have a
clear interpretation as being calibration conditions, in the generalised sense, for a certain type of D-branes in the geometry. This interpretation turns out to provide great tools to understand the geometry and discuss the stability of both supersymmetric and non-supersymmetric backgrounds.

\medskip

Generalised calibrations  are natural extensions of ordinary calibrations.\footnote{A calibration form $\omega$ is a $p$-form on $\mathcal{M}$ that satisfies an algebraic and a differential condition. 
At every point $q\in M$ and for every $p$-dimensional oriented subspace $\tau$
of the tangent space $T_q \mathcal{M}$
\begin{equation}
\label{pk:calbound}
 \omega|_{\tau} \leq \sqrt{\det g|_{\tau}} \, \dd \tau\equiv \text{vol}_\tau \,  ,
\end{equation}
where $\dd \tau = t^1 \wedge \ldots \wedge t^p$, with $t^\alpha$ a basis for $\tau^*$ (the dual of $\tau$) and $\det g|_{\tau}:=\det (g_{\alpha\beta})$, with $g_{\alpha\beta}$ the components of the pulled-back metric $g|_\tau$ in the coframe $t^\alpha$.
At every point there must exist subspaces $\tau$ such
that the above bound is saturated. 
Then the form $\omega$ must be closed  $\dd \omega=0$.}
For compactification on a special holonomy manifold $\mathcal{M}$, where there are no non-trivial bulk and world-volume fluxes, there is a nice 
relation between branes wrapping cycles in $\mathcal{M}$ and ordinary calibrations.
The calibration forms can be built as bilinears in the covariantly constant spinors on the manifold. As these spinors are the internal supersymmetry parameters,  the closure of the calibration form 
follows supersymmetry.
In this case, 
the energy of a brane wrapping a cycle in the manifold is given by its volume. Supersymmetric configurations are energy mininimising, and therefore correspond to branes wrapping calibrated cycles in the special holonomy manifold.

In flux compactifications, the energy of the  static branes gets contribution both from the volume and the fluxes. Generalised Complex Geometry provides a natural extension of this construction to flux backgrounds, which takes into account the contribution to the energy of both RR background fluxes and world-volume degrees of freedom \cite{Martucci:2005ht}.

\medskip

We consider D-branes in the warped geometry \eqref{10dmet}.
They can wrap a cycle $\Sigma$ in the internal manifold $\mathcal{M}$ and they can be string, domain-walls or space-filling in the external Minkowski space. As discussed in  \cite{Martucci:2005ht},  one can show that a static brane  wrapping a cycle $\Sigma$ in the internal manifold of an
$\mathcal{N}=1$  warped flux backgrounds is supersymmetric if it wraps a calibrated generalised submanifold.
 
To make these statements precise, we need to introduce the technology required to describe D-branes in generalised geometry: a generalised submanifold $(\Sigma,\ \mathcal{F})$ and a generalised calibration form $\omega$.

\medskip

A generalised submanifold is a pair $(\Sigma,\ \mathcal{F})$ with $\Sigma\subset\mathcal{M}$ a submanifold and $\mathcal{F}$ a two-form, which for a D-brane is a two-form on its world-volume, such that 
 \be 
 \dd \mathcal{F}=H|_\Sigma,\label{Hint}
 \ee
 with $H|_\Sigma$ the pullback of the NSNS-field-strength on $\Sigma$. The generalised submanifold is a generalised cycle if $\partial\Sigma=\emptyset$. 

\medskip

As shown in \cite{Martucci:2005ht}, one can construct  polyforms of definite parity  in terms of the pure spinors defining the SU$(3) \times$ SU$(3)$ structure of the $\mathcal{N}=1$ background
 \begin{align}
 \label{caldef}
  \omega^{\nn{string}}=&\ e^{2A-\phi}\text{Im}\Psi_1\\
 \omega^\nn{DW}=&\ e^{3A-\phi}\Psi_2\\
     \omega^{\nn{sf}}=&\ e^{4A-\phi}\text{Re}\Psi_1,
 \end{align}
 which satisfy the properties of a generalised calibration. They first satisfy an algebraic condition corresponding to the minimisation of the D-brane energy
\begin{align}
    \mathcal{E}(\Pi,\mathcal{R})\geq\ (\omega\wedge e^\mathcal{R})|_\Pi\label{bound}
    \end{align}
for any point $p\subset\mathcal{M}$ and any generalised submanifold $(\Pi, \mathcal{R})$\footnote{Strictly speaking for any point $p\subset\mathcal{M}$ there must exist a generalised submanifold $(\Pi, \mathcal{R})$ such that the above bound is saturated.}, with
 \be\mathcal{E}(\Pi,\mathcal{R})=e^{qA-\phi}\sqrt{\nn{det}(g|_\Pi+\mathcal{R})}-\delta_{q-1,3}e^{4A}\Tilde{C}|_\Pi,\ee
 where $q$ is the number of external dimensions and $\Tilde{C}$ the RR potentials such that 
 $\dd_H\Tilde{C}=\Tilde{F}$. Moreover the differential conditions that must be respected by the above generalised calibration forms correspond to the supersymmetry conditions \eqref{ret}, \eqref{second}, and \eqref{susy3}
\begin{align}
  \dd_H(\omega^\nn{DW})=\ 0&\qquad\text{domain-wall BPSness}\\ 
  \dd_H(\omega^\nn{string})=\ 0&\qquad\text{D-string BPSness}\\
   \dd_H(\omega^\nn{sf})=\ e^{4A}\Tilde{F}&\qquad\text{gauge BPSness}.
\end{align}
%These are exactly the pure spinor equations required to preserve %\mathcal{N}=1$ supersymmetry, which is therefore equivalent to the calibration, or BPSness, of the would-be domain-wall, space-filling and string-like D-branes. 

\medskip

A calibrated generalised cycle is a generalised cycle saturating the calibration bound \eqref{bound}. A D-brane in a $\mathcal{N}=1$ backgrounds is supersymmetric, or BPS, if it wraps a calibrated generalised cycle.
This is why we refer to the $\mathcal{N}=1$ supersymmetry conditions as the domain-wall, D-string and gauge BPSness respectively. The above generalised calibration forms are associated to space-filling, domain-wall, and string-like D-branes, which wrap respectively $4,\ 3$ and $2$ non-compact dimensions.

 \medskip

Another useful characterisation of D-branes is in terms of their generalised current. 

  The generalised current $j_{(\Sigma,\ \mathcal{F})}$ can be seen as the Poincaré dual of the generalised submanifold $(\Sigma,\ \mathcal{F})$:
        \be \int_\mathcal{M} \braket{\phi,j_{(\Sigma,\ \mathcal{F})}}=\int_\Sigma \phi|_\Sigma\wedge e^\mathcal{F}\label{currentdef}\ee
        with $\phi$ any polyform on $\mathcal{M}$. Loosely speaking, as a distribution $j_{(\Sigma,\ \mathcal{F})}$ is a localised real pure spinor proportional to $e^{-\mathcal{F}}\wedge \delta^{(d-k)}(\Sigma)$, with $\Sigma$ of rank $k$ and with $\delta^{(d-k)}(\Sigma)$ the standard Poincaré dual of the submanifold $\Sigma$. One can also consider the smeared version of this current, that we call $j$,  proportional to $e^{-\mathcal{F}}\wedge \nn{vol}_\perp$ with $\nn{vol}_\perp$ the transverse volume to $\Sigma$. 

        We can define the generalised tangent bundle of the foliation associated to the generalised submanifold $(\Sigma,\ \mathcal{F})$ as
        \begin{align}
T_{(\Sigma,\ \mathcal{F})}=&\{ V = v+\xi\in T\mathcal{M} \oplus T^\ast\mathcal{M} \, \ \big|\ \,  V \cdot j=0 \} \, \label{nullspace}\\
=&\{V=v+\xi\in T\Sigma\oplus T^\ast\mathcal{M} \, \  \big| \ \, \xi|_\Sigma=\iota_v\mathcal{F}\}.
\end{align}
This is a real maximally isotropic subbundle\footnote{A subbundle $L\subset E$ is maximal if dim$(L)=\nn{dim}(E)/2$ and isotropic if $(V,W)=0\quad\forall V,\ W\in L$.} of the generalised tangent bundle $E$.

As discussed in \cite{Martucci:2005ht}, the calibration condition \eqref{susy3} implies that $T_{(\Sigma,\ \mathcal{F})}$ is stable under the generalised complex structure $\mathcal{J}_2$ and hence  $(\Sigma, \mathcal{F})$ is a generalised complex submanifold. 

    Finally, it is easy to prove that the generalised current associated to a generalised cycle $(\Sigma,\ \mathcal{F})$ is $\dd_H$-closed. From \eqref{currentdef}, we have:
    \be \int_\mathcal{M} \braket{\phi,\dd_H j_{(\Sigma,\ \mathcal{F})}}=\int_\mathcal{M} \braket{\dd_H \phi,j_{(\Sigma,\ \mathcal{F})}}=\int_\Sigma \dd_H\phi|_\Sigma\wedge e^\mathcal{F}=\int_{\partial\Sigma}\phi|_{\partial\Sigma}\wedge e^{\mathcal{F}},\ee
    where we used both the property of the Mukai pairing \eqref{Mukaid} and Stoke theorem. Therefore we have
    \begin{equation}
        \dd_H j_{(\Sigma,\ \mathcal{F})}=j_{(\partial\Sigma,\mathcal{F}|_{\partial\Sigma})},
    \end{equation}
    which reduces to
      \begin{equation}
        \dd_H j_{(\Sigma,\ \mathcal{F})}=0
    \end{equation}
    if $(\Sigma,\ \mathcal{F})$ is a generalised cycle.
    
Note that, as the generalised tangent bundle $T_{(\Sigma, \mathcal{F})}$ is a real maximally isotropic sub-bundle of $T \mathcal{M} \oplus 
T^\ast \mathcal{M}$, it also defines an almost Dirac structure \cite{Gualtieri:2007ng}. From \eqref{nullspace} and Frobenius theorem it follows that $\dd_H j_{(\Sigma,\ \mathcal{F})}=0$ implies that the almost Dirac structure is actually integrable. 

\section{$\mathcal{N}=0$ flux vacua in Generalised Complex Geometry}\label{sec:n0gcg}

The goal of this paper is to construct and study new non-supersymmetric backgrounds. To do so we will focus on situations where supersymmetry breaking occurs as a small perturbation around some supersymmetric backgrounds  and it is  controlled by parameters whose vanishing would restore supersymmetry.

 Indeed, studying non-supersymmetric solutions of supergravity is hard, mainly because one has to deal with the full second order differential equations of motion. However, by restricting ourselves to these specific ways to break supersymmetry, it may be possible to keep some of the helpful properties of supersymmetric backgrounds, in particular their characterisation by first order 
 differential equations. 

The idea is to modify the Killing spinor equations \eqref{10dsusyg} and \eqref{10dsusyd}
while still assuming that the internal spinors  $\eta_1$ and $\eta_2$ in \eqref{10dspinors} are globally defined.  This means that the internal manifolds are still characterised by an  SU$(3)$ or SU$(2)$ structure, and, in the  Generalised Geometry language, by an SU$(3) \times$SU$(3)$ structure. 
The modified Killing spinor equations are then equivalent to adding supersymmetry breaking terms to the right-hand side of the  $\mathcal{N}=1$ pure spinor equations \eqref{ret}, \eqref{second}, and \eqref{susy3}.
From now on, we will call these new equations  modified pure spinor equations.

As discussed in the Section \ref{subsec:calib} the supersymmetry conditions \eqref{ret}, \eqref{second}, and \eqref{susy3} correspond to the calibration conditions for supersymmetric domain-wall, string-like and space-filling probe branes, respectively. Thus we will call the corresponding supersymmetry breaking terms, domain-wall (DWSB), string-like (SSB)  and space-filling supersymmetry breaking. 

DWSB non-supersymmetric backgrounds have been studied in the framework of GCG
in \cite{Lust:2008zd}, which also gave the general expression for the 
non-supersymmetric deformations of the Killing spinor equations and of the associated modified pure spinor equations.

In this section we will review a simple subclass of the DWSB discussed in \cite{Lust:2008zd} and we will discuss its geometrical properties, which we will then extend to our new classes of solutions describing SSB supersymmetry breaking, with and without DWSB breaking.

However, in contrast with the supersymmetric case, there is no reason to expect the solutions of the modified pure spinor equations and the Bianchi identities to be solutions of the equations of motion. So the supersymmetric breaking terms have to satisfy additional constraints in order to have a real vacuum, which we will discuss in Section \ref{sec:effpot}. 

\subsection{The DWSB vacua}
\label{subsec:DWSB}

Non-supersymmetric solutions corresponding to DWSB have been studied in \cite{Lust:2008zd}. The parametrisation of the most general DWSB deformation can be found in Appendix B of \cite{Lust:2008zd}. 
As it is hard to find solution in such general context, \cite{Lust:2008zd} focuses on a subset of solutions that only depend on a single supersymmetry breaking parameter. 

For the one-parameter DWSB class, the modified Killing spinor equations are\footnote{Here we write the modified dilatino variation defined in \eqref{moddilatino}.}
 \begin{align}
\delta\psi^{(1)}_\mu=\frac{1}{2}e^A \hat{\gamma}_\mu \zeta\otimes( r  \eta_1^\ast)
+c.c.\qquad&\delta\psi^{(2)}_\mu=\frac{1}{2}e^A \hat{\gamma}_\mu \zeta\otimes ( r \eta_2^\ast) +c.c.\\
\delta\psi^{(1)}_m=\zeta\otimes (-\frac{1}{2}r{\Lambda^n}_{m}\gamma_n\eta_1^\ast)+c.c.\qquad&\delta\psi^{(2)}_m=\zeta\otimes (-\frac{1}{2}r{\Lambda_m}^{n}\gamma_n\eta_2^\ast)+c.c.\\
\Delta\epsilon_1=\zeta\otimes (- r  \eta_1^\ast)  +c.c.\qquad&\Delta\epsilon_2=\zeta\otimes (- r  \eta_2^\ast) +c.c.
  \end{align}

As suggested by its name, the one-parameter DWSB subclass only  depends on a single  supersymmetry breaking parameter, $r$, as the O$(6)$ matrix  $\Lambda$ is completely defined by the background geometry:
\be 
\label{Umat}
\eta_1=iU\eta_2\qquad U\gamma_mU^{-1}={\Lambda^n}_{m}\gamma_n,
\ee
where $U$ is a unitary, point-dependent operator acting on six-dimensional spinors.

The corresponding domain-wall BPSness violation is then
\begin{align}
\label{modDW}
\dd_H(e^{3A-\phi}\Psi_2)=&ire^{3A-\phi}((-1)^{|\Psi_2|}\text{Im}\Psi_1+\frac{1}{2}\Lambda_{mn}\gamma^m\text{Im}\Psi_1\gamma^n).
\end{align}
where $\gamma^m$ are Cliff$(6)$ gamma matrices acting on a form $\omega$ as 
 \be 
 \label{gammas}
 \gamma^m\omega=(g^{mn}\iota_n+\dd y^m\wedge)\omega\qquad\omega\gamma^m=(-1)^{|\omega|+1}(g^{mn}\iota_n-\dd y^m\wedge)\omega.
 \ee

 \medskip
      
The modified domain-wall condition \eqref{modDW} can be rewritten in a way that makes explicit its implications for the geometry of the internal background \cite{Lust:2008zd}. 

We suppose that the internal manifold $\mathcal{M}$ admits a generalised submanifold $(\Sigma,\ \mathcal{F})$, where $\Sigma$ is a subbundle of odd/even dimension $n$ in type IIA/IIB. Since the space-filling calibration condition \eqref{susy3} still holds for this class of backgrounds, we can choose $(\Sigma, \mathcal{F})$ to be
calibrated by $\omega^{\nn{sf}}=\ e^{4A-\phi}\text{Re}\Psi_1$, such that the BPS space-filling branes of our backgrounds will wrap the calibrated generalised submanifold $(\Sigma, \mathcal{F})$.

We can then split the tangent bundle as 
\be
T \mathcal{M}=T \Sigma\oplus T \Sigma^\perp \, , 
\ee
with $ T \Sigma^\perp$ the orthogonal completion of $ T \Sigma$, and define a local vielbein $\{e^a\}$ on $T \Sigma\oplus T\Sigma^\perp$  and its associated gamma matrices
\be 
\hat{\gamma}^a\omega=(\delta^{ab}\iota_{b}+e^m\wedge)\omega\qquad\omega\hat{\gamma}^a=(-1)^{|\omega|+1}(\delta^{ab}\iota_{b}-e^a\wedge)\omega.\label{gammavielbein}
\ee
One can then express the operator $U$ in \eqref{Umat} as
\be
U = \gamma^n_{(6)} \sum_k \frac{\epsilon^{a_1 \ldots a_{n - 2k} b_1 \ldots b_{2k} }}{(n -2 k)! k! 2^k \sqrt{ \det (g|_\Sigma + \mathcal{F})}} \hat{\gamma}_{a_1 \ldots a_{n - 2k} } \mathcal{F}_{b_1 b-2} \ldots \mathcal{F}_{b_{2k- 1} b_{2k} },
\ee 
and the corresponding O$(6)$ matrix as
 \be \hat{\Lambda}=\mathbb{1}_\perp-(g|_\Sigma+\mathcal{F})^{-1}(g|_\Sigma-\mathcal{F}) \, ,
 \ee
where $\mathbb{1}_\perp$ is the projection onto $T\Sigma^\perp$.

Then \eqref{modDW} becomes 
\be  
\dd_H(\text{e}^{3A-\phi}\Psi_2)= irj,\label{jdwsb}
\ee
with
\be  j=4(-1)^{|\Psi_2|}e^{3A-\phi}\frac{\sqrt{\nn{det}g|_\Sigma}}{\sqrt{\nn{det}(g|_\Sigma+\mathcal{F})}}e^{-\mathcal{F}}\wedge\sigma
(\nn{vol}_\perp),
\ee
where $\nn{vol}_\perp$ is the volume form on the space orthogonal to the cycle $\Sigma$ 
such that $\nn{vol}_6=\nn{vol}_\Sigma\wedge\nn{vol}_\perp$, and $|\Psi_2|$ is the degree mod 2 of $\Psi_2$.   

The polyform $j$ is a smeared version of the Poincaré dual to the generalised cycle $(\Sigma,\ \mathcal{F})$, and therefore it is a (smeared) generalised current for $(\Sigma,\ \mathcal{F})$\footnote{ To illustrate, let's consider the simple cases where $\mathcal{F}=0$.
We have
    \be j=4(-1)^{|\Psi_2|}e^{3A-\phi}\sigma(\nn{vol}_\perp),\ee
which is the ordinary smeared Poincaré dual to the cycle $\Sigma$, and $T_{(\Sigma,0)}\equiv T\Sigma\oplus T^\ast\mathcal{M}|_{\Sigma^\perp}$ is the null space of $j$, which is a defining property of the generalised current. }. 
Moreover, the right-hand side of \eqref{jdwsb} is $\dd_H$-exact and so $\dd_H$-closed and, by  Frobenius's theorem,  it follows that the generalised sub-bundle 
$(\Sigma, \mathcal{F})$ is integrable and that 
$\dd \mathcal{F} = H|_\Sigma$.
This means that the manifold $\mathcal{M}$ is a 
foliation with leaves $(\Sigma, \mathcal{F})$, which are calibrated generalised submanifolds, thanks to \eqref{susy3}.

Note that solving \eqref{jdwsb}
for a given generalised foliation $(\Sigma,\ \mathcal{F})$ constrains the possible choices for $r$. Indeed, it has to be chosen such that the right-hand side of \eqref{jdwsb} is $\dd_H$ closed. The supersymmetry parameter can therefore not be multiplied by arbitrary complex functions, and these backgrounds truly depend on one parameter only.

 It is also important to note that the sources of these backgrounds are taken to be parallel, so their internal manifolds admit a unique generalised calibrated cycle $(\Sigma,\ \mathcal{F})$ wrapped by all the sources.
\medskip

The one-parameter DWSB class includes the GKP vacua \cite{Giddings:2001yu} as well as vacua with D4, D5 or D6-brane sources that can be obtained by T-dualising the GKP solution. 
Non-supersymmetric GKP vacua \cite{Giddings:2001yu} are solutions 
of type IIB compactifications, with D3-branes and O3-plane sources, and non trivial NSNS and RR three-form fluxes.

GKP-like vacua correspond to particularly simple representative of the one-parameter DWSB class of vacua, where we have
        \be \Lambda=  \mathbb{1} \ee 
        and $(\Sigma,\ \mathcal{F})$ is the trivial foliation whose leaves are points of $\mathcal{M}$.  In this case, the D-branes sitting on such leaves are D3-branes, and the failure to calibrate the would-be domain-wall branes originates purely from the $H_{(3,0)}$ components of the NSNS-flux\footnote{with respect to the almost complex structure defined by the internal spinor, for which $\Psi_2$ is a $(3,0)$-form.}:
\be
  \dd_H(\text{e}^{3A-\phi}\Psi_2)=\text{e}^{3A-\phi}H\wedge\Psi_2=4ir e^{3A-\phi}\nn{vol}_6. 
\ee

Backgrounds with D5 and D6-branes have been explicitly constructed in \cite{Lust:2008zd}, and we will revisit them when turning to the examples of our new backgrounds in Section \ref{sec:ex}.

\medskip

Let us consider again the generalised tangent bundle to the foliation with the generalised submanifold $(\Sigma, \mathcal{F})$. We saw that for supersymmetric backgrounds, $(\Sigma, \mathcal{F})$ is a complex generalised submanifold due to the integrability of the complex structure.
Thanks to this property it is possible to study 
deformation $(\Sigma, \mathcal{F})$ and then of 
D-branes in the background \cite{Martucci:2005ht}. 

For the one-parameter DWSB backgrounds $\mathcal{J}_2$ is not integrable anymore and one might wonder what can be said about $(\Sigma, \mathcal{F})$. 
Recall from Section \ref{subsec:calib} that the 
generalised tangent bundle $T_{(\Sigma,\ \mathcal{F})}$ associated to the foliation with the generalised submanifold $(\Sigma,\ \mathcal{F})$ is an almost Dirac structure. 
As pointed out in \cite{Lust:2008zd}, for 
DWSB vacua \eqref{jdwsb} implies the (conformal) closure of the associated generalised current and therefore
$T_{(\Sigma,\ \mathcal{F})}$  is still integrable: it remains closed under 
the twisted Courant bracket \eqref{twistcour}.

The integrability of the Dirac structure associated to $(\Sigma,\ \mathcal{F})$ has the important consequence that one can define a differential $\dd_{(\Sigma,\ \mathcal{F})}$ acting on the graded complex $\bigoplus_{k=0}^6 \Lambda^k T^\ast_{(\Sigma,\ \mathcal{F})}$.
It also allows to preserve some notion of "generalised holomorphicity". 
Let $L_2\subset E$ be the subbundle with $+i$-eigenvalues with respect to the generalised almost complex structure $\mathcal{J}_2$\footnote{$L_2$ is said to be the space of annihilators of $\Psi_2$, since $ V\cdot\Psi_2=0\ \forall\ V\in L_2$.}. $L_2$ defines an almost Dirac structure, which is not integrable as $\mathcal{J}_2$ is not. 
However, we can consider  the complex bundle
    \be L_{(\Sigma,\ \mathcal{F})}=T_{(\Sigma,\ \mathcal{F})}\cap L_2=\{V\in (T\oplus T^\ast)\otimes\mathbb{C}\ |\ V\cdot\Psi_2=V\cdot j=0\} \, , 
\ee
which, in contrast to $L_2$, is stable under the twisted Courant bracket:
    \begin{align}
        &[V,W]^H_C\cdot j=- W\cdot V\cdot\dd_Hj=0\\
          &[V,W]^H_C\cdot e^{3A-\phi}\Psi_2=-W\cdot V\cdot\dd_H(e^{3A-\phi}\Psi_2)=-ir W\cdot V\cdot j=0
    \end{align}
precisely because\footnote{Here we considered $e^{3A-\phi}\Psi_2$ for convenience but of course one finds the same result for $\Psi_2$.} of the modified pure spinor equation \eqref{jdwsb} and because of the integrability of the Dirac structure associated to $T_{(\Sigma,\ \mathcal{F})}$.
One can therefore define a differential that we call $\bar\partial_{(\Sigma,\ \mathcal{F})}$ on  $\bigoplus_{k=0}^3 \Lambda^k L^\ast_{(\Sigma,\ \mathcal{F})}$\footnote{The subbundle $ L_{(\Sigma,\ \mathcal{F})}$ is guaranteed to be three-dimensional from the compatibility of the pure spinors $\Psi_2$ and $j$, see for instance \cite{Gualtieri:2007ng}. }, and thus even though $\mathcal{J}_2$ is not integrable, the structure of the one-parameter DWSB class allows one to have a notion of holomorphic differential $\bar\partial_{(\Sigma,\ \mathcal{F})}$, at least with respect to the foliation with $(\Sigma,\ \mathcal{F})$.

It has been speculated in \cite{Lust:2008zd} that the first cohomology group of this differential $H^1_{\bar\partial_{(\Sigma,\ \mathcal{F})}}$ might define the moduli-space of the D-branes  in the one-parameter DWSB backgrounds, much like in the case of supersymmetric compactifications \cite{KoerberMartucci}. The authors also postulated that the non-integrability of the structure $\mathcal{J}_2$ might results in a closed string moduli space that is not a complex manifold, since it is what happens for the GKP construction. We refer the reader to \cite{Lust:2008zd} for more details on this.

\subsection{The SSB vacua}\label{subsec:SSB}

 In this section we will consider another class of non-supersymmetric backgrounds where supersymmetry is broken by deforming the D-string calibration condition \eqref{second}. We refer to this way of breaking supersymmetry as SSB, D-string or string-like supersymmetry breaking, to match \cite{Lust:2008zd}. 

The construction of such vacua relies on the same ingredient as for one-parameter DWSB solutions, namely the foliation structure of $\mathcal{M}$ by a generalised calibrated submanifold $(\Sigma,\ \mathcal{F})$.  

\medskip

An example of SSB solution has been discussed in type IIA \cite{Legramandi:2019ulq}. It is obtained by adding to a supersymmetric solution new components to the NSNS-field-strength, carefully chosen for the background to keep on respecting the equations of motion. The modified pure spinor equations read
\begin{align}
  \dd_H(\text{e}^{3A-\phi}\Psi_2)=&\ 0\\ 
  \dd_H(\text{e}^{2A-\phi}\text{Im}\Psi_1)=&\ e^{2A-\phi}H\wedge\text{Im}\Psi_1=c\ e^{6A-2\phi}\nn{vol}_6 \\
   \dd_H(e^{4A-\phi}\text{Re}\Psi_1)=&\ e^{4A}\Tilde{\ast}_6  F,
\end{align}
where c is a supersymmetry breaking parameter.
We refer the reader to \cite{Legramandi:2019ulq} for the explicit form of the solutions and the details of the construction.\footnote{Note that our pure spinors, volume form and NSNS-flux conventions differ to the ones in \cite{Legramandi:2019ulq}: $\Psi_\nn{ours}=-8i\Psi_\nn{theirs}$, $H_\nn{ours}=-H_\nn{theirs}$, $\nn{vol}_{6\ \nn{ours}}=-\nn{vol}_{6\ \nn{theirs}}$.}

It is important to note that the supersymmetry breaking term $c\ e^{6A-2\phi}\nn{vol}_6 $ has no given geometrical interpretation\footnote{The backgrounds considered in \cite{Legramandi:2019ulq} are intersecting NS5-D6-D8  models: the geometrical understanding of intersecting branes in GCG is fairly limited, and the literature on the subject is scarce, with the exception of \cite{Martucci4}.}, and then that no conclusion has been reached regarding the  stability of these backgrounds.

This is in stark contrast with the situation of the one-parameter DWSB class, where the supersymmetry breaking term $j$ is understood as the (smeared) generalised current associated to the background sources and is a key ingredient in the discussion of the stability of the one-parameter DWSB class.

It would therefore be interesting to construct non-supersymmetric backgrounds with string-like supersymmetry breaking, but this time with a supersymmetry breaking term that has a given geometrical interpretation.
\medskip

In this section, we will introduce a new class of SSB backgrounds, where the supersymmetry breaking term depends again on the (smeared) generalised current associated to the calibrated space-filling D-branes present in the background, much like the one-parameter DWSB case. The main motivation is that we can then benefit from the same kind of arguments when addressing the question of stability. 

Our starting point is the following assumption: the internal manifold admits a generalised foliation by 
the generalised cycle
$(\Sigma,\ \mathcal{F})$.
More precisely, we consider backgrounds that admit 
calibrated parallel space-filling sources and  we introduce a polyform $j$ which  plays the role of a (smeared) generalised current for our sources, which therefore wrap $(\Sigma,\ \mathcal{F})$, as in the DWSB case 
\be 
j=4(-1)^{|\Psi_2|}e^{3A-\phi}\frac{\sqrt{\nn{det}g|_\Sigma}}{\sqrt{\nn{det}(g|_\Sigma+\mathcal{F})}}e^{-\mathcal{F}}\wedge\sigma(\nn{vol}_\perp),
\ee
where $\nn{vol}_\perp$ is the transverse volume to the cycle $\Sigma$. 
The current $j$ is by construction $\dd_H$ closed as $(\Sigma,\ \mathcal{F})$ is a generalised cycle (see Section \ref{subsec:calib}).

The decomposition of the generalised current $j$ on the SU$(3)\times $SU$(3)$ structure is again
\be 
j=e^{3A-\phi}((-1)^{|\Psi_2|}\text{Im}\Psi_1+\frac{1}{2}\Lambda_{mn}\gamma^m\text{Im}\Psi_1\gamma^n)\label{jsu3su3}
\ee
with, as in the previous section, 
\be \hat{\Lambda}=\mathbb{1}_\perp-(g|_\Sigma+\mathcal{F})^{-1}(g|_\Sigma-\mathcal{F}) \, .
 \ee
However, in contrast with the DWSB construction, the domain-wall and the gauge BPSness conditions are both obeyed
 \begin{align}
  \dd_H(\text{e}^{3A-\phi}\Psi_2)=&\ 0 \label{retSSB}\\ 
   \dd_H(e^{4A-\phi}\text{Re}\Psi_1)=&\ e^{4A}\Tilde{\ast}_6  F,\label{calibsourcesssb}
\end{align}
and we consider the following violation of the D-string BPSness\footnote{ It is also useful to consider a local formulation in terms of the vielbein basis, which we will use when discussing concrete constructions of backgrounds:
\begin{align} 
\dd_H(\text{e}^{2A-\phi}\text{Im}\Psi_1)=&\ \hat{\alpha}_a[\hat{\gamma}^aj+(-1)^{|j|}j\hat{\gamma}^a]\label{SSB}\\
=&\ 2\hat{\alpha}_a e^a\wedge j.\label{SSBvielbein}\end{align}}
\begin{align} \dd_H(\text{e}^{2A-\phi}\text{Im}\Psi_1)=&\ \alpha_m[\gamma^mj+(-1)^{|j|}j\gamma^m]\label{SSBcoord}\\
=&\ 2\alpha_m\dd y^m\wedge j \, ,\label{SSBcoord2}
\end{align}
where the gamma matrices are defined in \eqref{gammas} and $\dd y^m$ span the directions 
transverse to the covolume of $\Sigma$. The coefficients  
$\{\alpha_m\}$ are real and are the supersymmetry breaking parameters. 

As for DWSB, imposing that the manifold is a generalised foliation
constrains the possible choices for the $\alpha_m$. Indeed, it
has to be chosen such that the right-hand side of \eqref{SSBcoord2} is $\dd_H$ closed. Therefore the $\alpha_m$ can't be multiplied by arbitrary complex functions, and these
backgrounds depend on dim$(\Sigma)$ parameters only.

The supersymmetry breaking term in \eqref{SSBcoord} can also be expanded on the SU$(3)\times $SU$(3)$ structure defined by the two pure spinors $\Psi_1$ and $\Psi_2$, it is given in \eqref{ssbdiamond}. As it is not a particularly insightful expression we will keep on using the parametrisation in terms of the generalised current from then on.

Note also that this ansatz is simply one of the different possibilities to write down an SSB term in terms of the generalised current. Our choice is dictated by simplicity: it is natural to construct backgrounds with such a supersymmetry breaking term and it  requires only relatively reasonable additional constraints in order to find solutions to the equations of motion, as we will discuss at length in Sections \ref{sec:effpot} and \ref{sec:ex}.

\medskip 

As for the one-parameter DWSB, 
the closure of the generalised current $j$ implies that 
the generalised bundle $T_{(\Sigma,\ \mathcal{F})}$ associated to the generalised submanifold $(\Sigma, \mathcal{F})$ defines a Dirac structure which is integrable. Then, it is again possible
 to define the differential $\dd_{(\Sigma,\ \mathcal{F})}$ acting on the graded complex $\bigoplus_{k=0}^6 \Lambda^k T^\ast_{(\Sigma,\ \mathcal{F})}$.

However, in contrast with the one-parameter DWSB situation, the generalised almost complex structure $\mathcal{J}_2$ is integrable, because of the conformal closure of the pure spinor $\Psi_2$ \eqref{retSSB}. Therefore, one can 
define a differential $\bar\partial_{\mathcal{J}_2}$ acting on the graded complex  $\bigoplus_{k=0}^6 \Lambda^k L^\ast_2$, where $L_2$ is the $+i$ eigenbundle of $\mathcal{J}_2$. 
The differential $\bar\partial_{\mathcal{J}_2}$ is the generalised Dolbeaut differential, see e.g \cite{Gualtieri:2007ng}.

Moreover, as the foliation $(\Sigma,\ \mathcal{F})$ is an almost generalised complex foliation, we can consider the three-dimensional complex sub-bundle
    \be L_{(\Sigma,\ \mathcal{F})}=T_{(\Sigma,\ \mathcal{F})}\cap L_2=\{V\in (T\oplus T^\ast)\otimes\mathbb{C}\ |\ V\cdot\Psi_2=V\cdot j=0\} \, , 
    \ee
which is stable under the twisted Courant bracket:
    \begin{align}
        &[V,W]^H_C\cdot j=- W\cdot V\cdot\dd_Hj=0\\
          &[V,W]^H_C\cdot e^{3A-\phi}\Psi_2=-W\cdot V\cdot\dd_H(e^{3A-\phi}\Psi_2)=0 \, . 
    \end{align}
%thanks to the facts that $j$ is the generalised current of the generalised cycle $(\Sigma,\ \mathcal{F})$ and that the generalised complex structure $\mathcal{J}_2$ is integrable.

As in the one-parameter DWSB case, one can therefore define a differential that we call $\bar\partial_{(\Sigma,\ \mathcal{F})}$ on the graded complex  $\bigoplus_{k=0}^3 \Lambda^k L^\ast_{(\Sigma,\ \mathcal{F})}$. It is then reasonable to postulate that the first cohomology group of this differential $H^1_{\bar\partial_{(\Sigma,\ \mathcal{F})}}$ might define the moduli-space of the D-branes present in our backgrounds, like in the supersymmetric  case. 
Note that  the complex bundle $L_{(\Sigma,\ \mathcal{F})}$ and thus the cohomology group $H^1_{\bar\partial_{(\Sigma,\ \mathcal{F})}}$ clearly depend only on one of the pure spinors, $\Psi_2$. This fact has a clear interpretation for supersymmetric backgrounds, when described from the four-dimensional perspective, see \cite{KoerberMartucci}, whereas we do not have access to an analogous explanation, since we have much less understanding of the four-dimensional theories associated to our backgrounds (see Section \ref{sec:effpot}).
    
    Moreover, the fact that this way of breaking supersymmetry preserves the integrability of the generalised complex structure $\mathcal{J}_2$ seems to suggest that the closed string moduli space could still be a complex manifold. It could be useful to turn to the Exceptional Generalised Geometry formalism \cite{Waldram2} to address these questions, for instance by performing a non-supersymmetric analogue analysis of the one developed in \cite{Waldram}. However, making precise statements about these problems is beyond the scope of the present work.

\medskip

Finally, from now on, we consider generalised foliations with $\mathcal{F}=0$. The form degree of the supersymmetry breaking term is therefore codim$(\Sigma)+1$. Then \eqref{SSB} can only be respected if
\be \text{type}(\Psi_1)\leq\text{codim}(\Sigma),
\ee
where the type of a pure spinor is its lower form degree. For instance backgrounds with an SU$(3)$ structure in type IIA have $\text{type}(\Psi_1)=3$ so dim$(\Sigma)\leq 3$, which corresponds to D$4$ or D$6$ branes only. We will explicitly construct some backgrounds with such sources in Section \ref{sec:ex}.

\subsection{Vacua with both DWSB and SSB contributions}\label{subsec:ssbdwsb}

The two classes of symmetry breaking  described 
so far correspond each to the failure to respect one specific differential calibration condition, either the domain-wall or the string-one.  
Here, we would like to discuss more general cases  where supersymmetry is broken by violating both the domain-wall and string-like calibration condition, while keeping  calibrated space-filling sources in the background.

In the light of the previous discussion, it is  natural to consider non-supersymmetric backgrounds which combine  the specific DWSB and SSB contributions discussed in the previous sections
 \begin{align}
  \dd_H(\text{e}^{3A-\phi}\Psi_2)=&\ irj\label{rssbdwsb}\\ 
  \dd_H(\text{e}^{2A-\phi}\text{Im}\Psi_1)=&\ \alpha_m[\gamma^mj+(-1)^{|j|}j\gamma^m]\label{alphassbdwsb}\\
   \dd_H(e^{4A-\phi}\text{Re}\Psi_1)=&\ e^{4A}\Tilde{\ast}_6  F.\label{RRdwsbssb}
\end{align}
Here again the internal space is taken to be a generalised foliation $(\Sigma,\ \mathcal{F})$, and $j$ is its smeared generalised current, while the sources wrap the calibrated generalised cycle $\Sigma$.

    The main geometric properties of the one-parameter DWSB class will be preserved by the additional SSB contribution: the subbundle $T_{(\Sigma,\ \mathcal{F})}$ also defines an integrable Dirac structure, the generalised complex structure $\mathcal{J}_2$ is not integrable, and the foliation $(\Sigma,\ \mathcal{F})$ is an almost generalised complex foliation, so one can still define the following complex bundle
     \be L_{(\Sigma,\ \mathcal{F})}=T_{(\Sigma,\ \mathcal{F})}\cap L_2=\{V\in (T\oplus T^\ast)\otimes\mathbb{C}\ |\ V\cdot\Psi_2=V\cdot j=0\},\ee which is stable under the Courant bracket. One can therefore again define the differential $\bar\partial_{(\Sigma,\ \mathcal{F})}$ on  $\bigoplus_{k=0}^3 \Lambda^k L^\ast_{(\Sigma,\ \mathcal{F})}$, which provides a notion of a holomorphic differential, at least with respect to $(\Sigma,\ \mathcal{F})$.
    
        We will come back to this class of vacua later on, write down its equations of motion in Section \ref{sec:effpot} and construct some explicit backgrounds solving these equations in  Section \ref{sec:ex}. 

\section{Effective potential and equations of motion from pure spinors}\label{sec:effpot}

In the previous sections, we presented different classes of non-supersymmetric backgrounds in terms of the modified pure spinor equations they obey. However, unlike the supersymmetric case, the solutions of the modified pure spinor equations (plus Bianchi identities) are not guaranteed to solve the full set of
equations of motion and thus to describe true vacua of type II supergravity. 

In order to check that the solutions of the modified pure spinor equations are actual supergravity backgrounds, we will follow the strategy of \cite{Lust:2008zd}: write the most general four-dimensional `effective potential' from the ten-dimensional type II supergravity action, and use the fact that the extremisation of the four-dimensional action is equivalent to satisfying the ten-dimensional type II supergravity equations of motion. 

The term `effective potential' is a bit misleading since it does not come from a truncation of the ten-dimensional theory to a finite set of four-dimensional modes. 

However, the reason behind this choice  (instead of just tackling the ten-dimensional equations of motion upfront) is two-fold. First of all, the effective potential can be written as an integral over the internal space of expressions involving the pure spinors and, by varying it, one can express the equations of motion as some tractable differential equations on the pure spinors. Secondly, the effective potential allows us to discuss the potential presence of closed string tachyons from the four-dimensional perspective and thus (partially) address the question of stability of our different backgrounds. For our Minkowski backgrounds, this translates into the requirement that the effective potential must be positive semi-definite. 

 Indeed, even though we do not perform a complete reductions to the four-dimensional effective theories, we will be able to find natural `truncations' suggested by the geometry for our new classes of backgrounds such that the effective potential will be positive semi-definite, therefore excluding the potential presence of closed string tachyons. The semi-definite positiveness will be argued through the use of calibration bounds such as \eqref{bound} rewritten in the pure spinor formalism.
\medskip

The equations of motion in terms of pure spinors are hard to obtain in full generality, and they have not been derived yet.
However, they have been written down in \cite{Lust:2008zd} for the case of the one-parameter DWSB family, using the specific forms of the modified pure spinor equations discussed in Subsection \ref{subsec:DWSB}. In these cases it has been shown that the equations of motion reduce to relatively mild additional constraints to add on top of the modified pure spinor equations.

\medskip

We start this section by recalling the general expression of the effective potential in terms of the pure spinors  derived in \cite{Lust:2008zd}. 
In Subsection \ref{subsec:veffeomdwsb} 
we summarise the results for the one-parameter DWSB case, presenting its effective potential and equations of motion. Subsections \ref{subsec:SSB} and \ref{subsec:ssbdwsb}
contain our new results on the effective potential and equations of motion for both purely 
 SSB constructions and mixed SSB and DWSB constructions. We first derive the equations of motion for the most general violation of the D-string calibration condition, and we then specify them to our two new classes of backgrounds.

\subsection{The type II effective potential from pure spinors}\label{subsec:veff}

In this section, we recall the derivation of
the `effective' potential of  \cite{Lust:2008zd}.
We are interested in configurations where the space-time is warped 
\be 
\dd s^2_{10}=e^{2A(y)}g_{\mu\nu}\dd x^\mu\dd x^\nu+g_{mn}\dd y^m\dd y^n,
\ee
with $g_{\mu\nu}$ now a generic four-dimensional metric, and with non-trivial NSNS and RR-fluxes. We assume that the metric $g_{\mu\nu}$  depends only on the external coordinates, while all the other fields, warp factor, internal metric and fluxes 
depend only on the internal coordinates.

The effective four-dimensional action for such configurations takes the form
\be 
S_{\text{eff}}=\int_{X_4} \dd^4x\sqrt{-g_4}\left(\frac{1}{2}\mathcal{N}R_4-2\pi\mathcal{V}_{\text{eff}}\right),\label{seff}
\ee
where $R_4$ is the four-dimensional scalar curvature, $\mathcal{N}$ is the warped volume of the internal space
\be 
\mathcal{N}=4\pi \int_{\mathcal{M}}e^{2A-2\phi}\ \text{vol}_6
\ee
and the effective potential density is given by 
\begin{align}
\label{effpot1}
 \mathcal{V}_{\nn{eff}}=&\int_{\mathcal{M}}\nn{vol}_6 e^{4A}\{e^{-2\phi}[-\mathcal{R}+\frac{1}{2}H^2-4(\dd\phi)^2+8\nabla^2A+20(\dd A)^2]-\frac{1}{2}\Tilde{F}^2\}\nonumber\\
        &+\sum_{i \in \nn{loc. sources}}\tau_i\left(\int_{\Sigma_i}e^{4A-\phi}\sqrt{\nn{det}(g|_{\Sigma_i}+\mathcal{F}_i)} -\int_{\Sigma_i} \tilde{C}|_{\Sigma_i}\wedge e^{\mathcal{F}_i}\right) \, ,
    \end{align} 
with  $\mathcal{R}$ the six-dimensional scalar curvature.     

The first line in $\mathcal{V}_{\nn{eff}}$ corresponds to the closed string sector, while the second line is the contribution from localised sources. 
We follow the conventions of \cite{Lust:2008zd} where 
$2\pi\sqrt{\alpha'}=1$,  so that the tensions of all D-branes are equal $\tau_{\nn{D}_p}=1$ and for 
O-planes $\tau_{O_q}=-2^{q-5}$.
Notice that  we are also omitting the internal field kinetic terms, since they are taken to be constant along the external directions\footnote{As in 
\cite{Lust:2008zd} we neglect anomalous curvature-like corrections to the sources
contribution and, for O-planes, we take $\mathcal{F}=0$, as they are not seen
as dynamical objects in the compactification.}.
 The sources couple to the RR potentials defined by $\dd_H \tilde{C} =e^{4A}\Tilde{F}$.

As argued in \cite{Lust:2008zd},  the variations of the four dimensional action \eqref{seff} exactly reproduce the ten-dimensional equations of motion (see  Appendix \ref{sec:AppA} for the expression of the equations of motion). Moreover, from the variation of the four-dimensional action with respect to $g_{\mu\nu}$\footnote{which from the ten-dimensional perspective is equivalent to the internal space integral of the external ten-dimensional Einstein equation's trace.}, one gets that the external space is Einstein, with \be R_4=8\pi\mathcal{V}_\text{eff}/\mathcal{N}.\ee

Notice also that the variation of the effective action with respect to the electric RR potentials  reproduces the Bianchi identities
\be 
\dd_HF=-j_\text{tot}=-\sum_{i}\tau_ij_i,\label{bianchi}
\ee
where, as described in  Section \ref{subsec:calib}, $j_i$ are the (smeared) generalised current associated to the D-branes wrapping cycles on the internal manifold. 

In the following discussions it will be convenient to consider the external component of the modified Einstein equations \eqref{modeinstein}. This can be obtained by combined variations of  \eqref{seff} 
\be 
\frac{\delta S_\nn{eff}}{\delta A}+2 \frac{\delta S_\nn{eff}}{\delta \phi}=0.\label{modeinstein2}
\ee
        %Since we will impose the vanishing of the variation of the effective action with respect to the dilaton, which amounts to the dilaton equation of motion, we therefore see from \eqref{modeinstein2} that imposing the vanishing of the variation of the effective action with respect to the warp factor is equivalent to the external component of the modified Einstein equations.
A central result of \cite{Lust:2008zd} is the fact that the effective potential \eqref{effpot1} can be expressed in term of the pure spinors. The general expression is 
\begin{align}
\label{genVeff}
 \mathcal{V}_\text{eff}=&
\frac{1}{2}\int_{\mathcal{M}}  \braket{\Tilde{\ast}_6 [\dd_H(\text{e}^{2A-\phi}\text{Im}\Psi_1)],\dd_H(\text{e}^{2A-\phi}\text{Im}\Psi_1)}\nonumber \\
&+ \frac{1}{2}\int_{\mathcal{M}} e^{-2A}    \braket{\Tilde{\ast}_6 [\dd_H(\text{e}^{3A-\phi}\Psi_2)],\dd_H(\text{e}^{3A-\phi}\bar{\Psi}_2)}\nonumber\\
&+\frac{1}{2}\int_{\mathcal{M}}\text{vol}_6 \ e^{4A}|\Tilde{\ast}_6  F- e^{-4A}\dd_H(e^{4A-\phi}\text{Re}\Psi_1)|^2\nonumber\\
&-\frac{1}{4}\int_{\mathcal{M}}e^{-2A}\left(\frac{|\braket{\Psi_1,\dd_H(\text{e}^{3A-\phi}\Psi_2)}|^2}{\text{vol}_6}+\frac{|\braket{\bar{\Psi}_1,\dd_H(\text{e}^{3A-\phi}\Psi_2)}|^2}{\text{vol}_6}\right)\nonumber\\
    &-4\int_{\mathcal{M}}\text{vol}_6 e^{4A-2\phi}[(u^1_R)^2+(u^2_R)^2] \nonumber\\
     &+\sum_{i\subset\text{D-branes}}\tau_i \int_{\mathcal{M}}e^{4A-\phi}(\text{vol}_6 \ \rho_i^\text{loc}-\braket{\text{Re}\Psi_1, j_i}) \nonumber\\
&+\int_{\mathcal{M}}\braket{e^{4A-\phi}\text{Re}\Psi_1-\Tilde{C},\dd_H F+j_\text{tot}} \, ,
    \end{align}
where the square of a polyform is defined in Appendix \ref{sec:AppA}, and
\be u^{1,2}_R= u^{1,2}_{Rm}\dd y^m\equiv(u^{1,2}_m+u^{\ast 1,2}_m)\dd y^m \, ,
\ee    
with
\begin{align}
u^1_m=&\frac{i\braket{\gamma_m\bar{\Psi}_1,\dd_H(\text{e}^{2A-\phi}\text{Im}\Psi_1)}}{e^{2A-\phi}\braket{\Psi_1,\bar{\Psi}_1}}+\frac{\braket{\gamma_m\bar{\Psi}_2,\dd_H(\text{e}^{3A-\phi}\Psi_2)}}{2e^{3A-\phi}\braket{\Psi_2,\bar{\Psi}_2}}\\
u^2_m=&\frac{i(-1)^{|\Psi_2|}\braket{\Psi_1\gamma_m,\dd_H(\text{e}^{2A-\phi}\text{Im}\Psi_1)}}{e^{2A-\phi}\braket{\Psi_1,\bar{\Psi}_1}}+\frac{(-1)^{|\Psi_1|}\braket{\bar{\Psi}_2\gamma_m,\dd_H(\text{e}^{3A-\phi}\Psi_2)}}{2e^{3A-\phi}\braket{\Psi_2,\bar{\Psi}_2}}.
    \end{align}
We have also introduced the Born-Infeld density $\rho^\nn{loc}_i$ associated with a source wrapping a generalised submanifold $(\Sigma_i,\mathcal{F}_i)$
    \be \rho_i^\nn{loc}=\frac{\sqrt{\nn{det}(g|_{\Sigma_i}+\mathcal{F}_i)}}{\sqrt{\nn{det}g}}\delta(\Sigma_i).\ee

With this definition, it's insightful to rewrite the algebraic inequality \eqref{bound} in terms of $\rho^\nn{loc}$:
\be \rho^\nn{loc}_i\geq\frac{\braket{\nn{Re}\Psi_1,j_i}}{\text{vol}_6},\label{boundrho}
\ee
where the division by the volume form means that we remove the vol$_6$ factor in the numerator.
We take the sources to be calibrated as boundary conditions, which corresponds to the saturation of this bound. This bound also implies that the sixth line in the expression 
\eqref{genVeff}  of 
$\mathcal{V}_\nn{eff}$ is always positive.

\medskip

Varying the effective potential 
\eqref{genVeff} with respect to the dilaton, the NSNS and RR-fields, and the warp factor, one should obtain the ten-dimensional equations motion directly in terms of the pure spinors. However, so far this has not been done in the general case.

In the following, we derive the equations of motion for the most general violation of the D-string calibration, and we specify them for the  cases of our DWSB and SSB constructions presented in Subsections \ref{subsec:SSB} and \ref{subsec:ssbdwsb}.

\subsection{DWSB effective potential and equations of motion}\label{subsec:veffeomdwsb}

We can now turn to the case of the one-parameter DWSB construction, and extremise its effective potential to find out what are the additional constraints to impose on the one-parameter DWSB solutions to promote them to actual supergravity vacua. We briefly present the results, since they have been derived in \cite{Lust:2008zd}.

\medskip

Given that the D-string and gauge BPSness conditions \eqref{second} and \eqref{susy3} 
are respected for the one-parameter DWSB class, the first and third line of the effective potential \eqref{genVeff} 
%\eqref{dstring} and \eqref{gauge} 
vanish.  Then, combining D-string BPSness \eqref{second} and the modified DWSB condition \eqref{jdwsb} one can show that the terms proportional to  $u^1$ and $u^2$  in \eqref{genVeff} are also
zero. 
Finally, the saturation of the calibration bound \eqref{boundrho} and the Bianchi identities \eqref{bianchi} imply that the last two lines of \eqref{genVeff} vanish, leaving 
     \begin{align}
        \mathcal{V}_\text{eff}=& \frac{1}{2}\int_{\mathcal{M}} e^{-2A}    \braket{\Tilde{\ast}_6 [\dd_H(\text{e}^{3A-\phi}\Psi_2)],\dd_H(\text{e}^{3A-\phi}\bar{\Psi}_2)}\nonumber\\
    &-\frac{1}{4}\int_{\mathcal{M}}e^{-2A}\left(\frac{|\braket{\Psi_1,\dd_H(\text{e}^{3A-\phi}\Psi_2)}|^2}{\text{vol}_6}+\frac{|\braket{\bar{\Psi}_1,\dd_H(\text{e}^{3A-\phi}\Psi_2)}|^2}{\text{vol}_6}\right).\label{veffdwsb}
    \end{align}

The effective potential \eqref{veffdwsb} is sufficiently simple to derive the corresponding equations of motion \cite{Lust:2008zd}. We simply give the main results of the analysis and we refer to \cite{Lust:2008zd} for more details. $\mathcal{V}_\text{eff}$ depends explicitly on the warp factor, the dilaton and the NSNS-field-strength, and it depends implicitly on the internal metric through the Hodge operator, the pure spinors and the volume form. 
One should therefore vary it  with respect to these fields to find the equations of motions.  

The dilaton equation is obtained by varying $\mathcal{V}_\text{eff}$ with respect to the dilaton and it 
is satisfied identically by imposing the modified domain-wall BPSness equation \eqref{jdwsb}.

As we will also discuss in the next subsection, the external modified Einstein equation 
can be written as
        \begin{align}
        \braket{e^{4A}\Tilde{F}-\dd_H (e^{4A-\phi}\nn{Re}\Psi_1),F+\dd_H\Tilde{\ast}_6  (e^{-\phi}\nn{Re}\Psi_1)}&\nonumber\\
        +e^{4A-\phi}\sum_{i \in \nn{loc. sources}}\tau_i\Big[\rho_i^\nn{loc}\nn{vol}_6-\braket{\nn{Re}\Psi_1,j_i}\Big]&=0,\label{MEEEQ}
        \end{align}
which is identically satisfied when the gauge BPSness is obeyed and the calibration bound \eqref{boundrho} is saturated. 
The  external components of the modified Einstein equations are satisfied not only for the one-parameter DWSB class, but for any background preserving the D-string and gauge BPSness and violating the domain-wall calibration condition.

For the internal Einstein equation, the variation 
of the effective potential with respect to the internal metric gives 
\be 
\text{Im}\Big\{\braket{g_{k(m}\dd y^k\wedge\iota_{n)}\Psi_2,\dd_H\big[e^{A-\phi}r^\ast(3\nn{Re}\Psi_1+\frac{1}{2}(-1)^{|\Psi_2|}\Lambda_{pq}\gamma^p\nn{Re}\Psi_1\gamma^q)\big]}\Big\}=0,
\ee
which imposes some non-trivial constraints the  DWSB configurations must satisfy to be true supergravity solutions. 

Another set of constraints comes from the  variation of $\mathcal{V}_\text{eff}$ with respect to the NSNS-field 
\be \dd\Big[e^{4A-2\phi}\braket{\nn{Im}(r^\ast\Psi_2),3\nn{Re}\Psi_1+\frac{1}{2}(-1)^{|\Psi_2|}\Lambda_{mn}\gamma^m\nn{Re}\Psi_1\gamma^n}_3\Big]=0 \, . 
\ee

Interestingly, the NSNS-field equation of motion and the internal Einstein equation can be unified into the following condition:
 \be \int_{\mathcal{M}}e^{A-\phi}\nn{Im}\big\{r^\ast\braket{\delta_{g,B}[\dd_H(e^{3A-\phi}\Psi_2)],3\nn{Re}\Psi_1+\frac{1}{2}(-1)^{|\Psi_2|}\Lambda_{mn}\gamma^m\nn{Re}\Psi_1\gamma^n}\big\}=0 \, , \label{integrbgdwsb}\ee
where $\delta_{g,B}$ denote a generic deformation  of internal metric and $B$-field.  Given that the one-parameter DWSB backgrounds satisfy
    \be \braket{\dd_H(e^{3A-\phi}\Psi_2),3\nn{Re}\Psi_1+\frac{1}{2}(-1)^{|\Psi_2|}\Lambda_{mn}\gamma^m\nn{Re}\Psi_1\gamma^n}=0,\label{noscaledwsb}\ee
the conditions \eqref{integrbgdwsb} can be seen as a stability condition of \eqref{noscaledwsb} under deformations of $\dd_H(e^{3A-\phi}\Psi_2)$, thus providing a four-dimensional interpretation of these equations of motion, in terms of a stability condition for some F-flatness condition under such deformations, see \cite{Lust:2008zd} for more details.

\medskip

Finally, note that the effective potential \eqref{veffdwsb} vanishes when evaluated on the one-parameter DWSB background. 
This can be seen by directly  expanding the effective potential \eqref{veffdwsb}
on the SU$(3)\times $SU$(3)$ structure: we find that the terms in the two lines compensate exactly, as expected for backgrounds with Minkowski$_4$ external spaces. 

Alternatively we can also use \eqref{jdwsb} to rewrite the effective potential in terms of the generalised current $j$ associated to the generalised foliation $(\Sigma,\ \mathcal{F})$
      \be  
      \mathcal{V}_\text{eff}= \frac{1}{2}\int_{\mathcal{M}} \text{e}^{-2A}|r|^2\Big[\braket{\Tilde{\ast}_6 j,j}-\frac{|\braket{\Psi_1,j}|^2}{\text{vol}_6}\Big] \, ,\label{veffdwsbj}
      \ee
which will turn out to be an insightful formulation when considering the off-shell one-parameter DWSB potential in Subsection \ref{subsec:stab}.

The vanishing of the one-parameter DWSB effective potential can here be interpreted as the saturation of a calibration bound. Indeed, if $\Tilde{\jmath}_{(\Pi, \mathcal{R})}$ is the generalised current of a submanifold $(\Pi, \mathcal{R})$, which is not necessarily calibrated, the following bound can be derived from  \eqref{bound}
       \be \braket{\Tilde{\ast}_6 \Tilde{\jmath}_{(\Pi, \mathcal{R})},\Tilde{\jmath}_{(\Pi, \mathcal{R})}}\geq\frac{|\braket{\Psi_1,\Tilde{\jmath}_{(\Pi, \mathcal{R})}}|^2}{\nn{vol}_6}.\label{boundj}\ee
This bound gets saturated if the generalised submanifold is calibrated, which is the case for the one-parameter DWSB backgrounds, and the effective potential therefore vanish.

\subsection{SSB effective potential and equations of motion}\label{subsec:veffeomssbdwsb}

We consider now the SSB configurations introduced in  Subsection \ref{subsec:SSB}. 
Since SSB backgrounds haven't been studied in details in the literature, we discuss this case in more details. 

\medskip

We consider the most general violation of the D-string BPSness, without any non-calibrated sources in the backgrounds. We also assume that the Bianchi identities are respected.

    We first discuss how the effective potential simplifies.
The domain-wall and gauge BPS conditions, \eqref{ret} and  \eqref{susy3}
set the second, third and fourth line in \eqref{genVeff} to zero. 
Also in this case, the last two lines of \eqref{genVeff} vanish due to
the saturation of the bound \eqref{boundrho}  and the Bianchi identities \eqref{bianchi}. 
Then  the  effective potential is 
\begin{align}
        \mathcal{V}_\text{eff}=&
         \frac{1}{2}\int_{\mathcal{M}}  \braket{\Tilde{\ast}_6 [\dd_H(\text{e}^{2A-\phi}\text{Im}\Psi_1)],\dd_H(\text{e}^{2A-\phi}\text{Im}\Psi_1)}\nonumber\\
    &-4\int_{\mathcal{M}}\text{vol}_6 e^{4A-2\phi}[(u^1_R)^2+(u^2_R)^2],
\end{align}
where $\{u^{1,2}_m\}$ reduce to
   \begin{align}
u^1_m=&\frac{i\braket{\gamma_m\bar{\Psi}_1,\dd_H(\text{e}^{2A-\phi}\text{Im}\Psi_1)}}{e^{2A-\phi}\braket{\Psi_1,\bar{\Psi}_1}}\\
u^2_m=&\frac{i(-1)^{|\Psi_2|}\braket{\Psi_1\gamma_m,\dd_H(\text{e}^{2A-\phi}\text{Im}\Psi_1)}}{e^{2A-\phi}\braket{\Psi_1,\bar{\Psi}_1}}.
    \end{align}

Varying $\mathcal{V}_\text{eff}$ 
with respect to the dilaton gives
the dilaton equation of motion
\be 
\braket{\dd_H(e^{2A-\phi}\nn{Im}\Psi_1),\Xi}=0,
\ee
while the variation with respect 
to NSNS-field $B$ is 
\be \dd\big[e^{2A-\phi}\braket{\nn{Im}\Psi_1,\Xi}_3\big]=0 \, .  
\ee
The polyform $\Xi$ is defined as 
    \be \Xi=\Tilde{\ast}_6 \dd_H(e^{2A-\phi}\nn{Im}\Psi_1)+2e^{2A-\phi}u^1_{Rm}\gamma^m\nn{Re}\Psi_1+2(-1)^{|\Psi_2|}e^{2A-\phi}u^2_{Rm}\nn{Re}\Psi_1\gamma^m.\ee

We are left with the Einstein equations. To derive  the internal component of the Einstein equations, one needs the following rules for the variations with respect to the internal metric
            \begin{align}
                \delta\sqrt{\nn{det}g}=&-\frac{1}{2}\delta g^{mn}g_{mn}\sqrt{\nn{det}g}\\
    \delta\braket{\Tilde{\ast}_6 \chi_1,\chi_2}=&\ \delta g^{mn}\big[\braket{\Tilde{\ast}_6 \iota_m\chi_1,\iota_n\chi_2}-\frac{1}{2}g_{mn}\braket{\Tilde{\ast}_6 \chi_1,\chi_2}\big]\\
    \delta\Psi_i=&-\frac{1}{2}\delta g^{mn}g_{k(m}\dd y^k\wedge\iota_{n)}\Psi_i\qquad i=1,2.
            \end{align}
Then, we find that the internal Einstein equations read
\be
\braket{g_{k(m}\dd y^k\wedge\iota_{n)}(e^{2A-\phi}\nn{Im}\Psi_1),\dd_H\Xi}-\braket{g_{k(m}\dd y^k\wedge\iota_{n)}\dd_H(e^{2A-\phi}\nn{Im}\Psi_1),\Xi}=0.
\ee

It would be interesting to further develop $\Xi$ by plugging in the general SU$(3)\times $SU$(3)$ decomposition of $\dd_H(e^{2A-\phi}\nn{Im}\Psi_1)$ with only D-string BPSness violation \eqref{genstring}. One could then use the fact that the $\Tilde{\ast}_6 $ operator eigenstates are also eigenstates of the SU$(3)\times $SU$(3)$ structure\footnote{See for instance \cite{Tomasiello}.} to write these equations of motion on the generalised Hodge diamond, as first order differential equations on the supersymmetry breaking parameters introduced in Appendix \ref{sec:AppC}. One could then search for more general non-supersymmetric solutions of type II supergravity with SSB-like supersymmetry breaking.
We won't do this here, as we will focus on our ansatz in \ref{subsec:SSB}.

To study the external component of the modified Einstein equation we will follow \cite{Lust:2008zd}.  We first reduce 
the ten-dimensional equation \eqref{modeinstein}  on our  warped configurations 
\be 
\nabla^m(e^{-2\phi}\nabla_me^{4A})=e^{4A}\Tilde{F}\cdot\Tilde{F}+e^{4A-\phi}\sum_{i \in \nn{loc. sources}}\tau_i\rho_i^\nn{loc},
\ee
and rewrite it in terms of pure spinors as         
\begin{align} 
\label{extmodE}
-\dd(e^{-2\phi}\ast_6\dd e^{4A})=&\braket{\Tilde{\ast}_6 \Tilde{F},e^{4A}\Tilde{F}}-\braket{\dd_H\Tilde{\ast}_6 \Tilde{F},e^{4A-\phi}\nn{Re}\Psi_1} \nonumber \\
&+e^{4A-\phi}\sum_{i \in \nn{loc. sources}}\tau_i\Big[\rho_i^\nn{loc}\nn{vol}_6-\braket{\nn{Re}\Psi_1,j_i}\Big]
\end{align}
by using the Bianchi identity \eqref{bianchi} together with the RR-field-strength self-duality \eqref{fselfdual}.

At this point we see the difference between  the DWSB configurations of \cite{Lust:2008zd}
and the SSB ones. For DWSB, using  D-string and gauge BPSness, one can prove the identity 
\be 
\dd(e^{-2\phi}\ast_6\dd e^{4A})=\dd\braket{\Tilde{\ast}_6 \dd_H (e^{4A-\phi}\nn{Re}\Psi_1),e^{-\phi}\nn{Re}\Psi_1}_5,\label{rewritingmeeeq}
\ee
which in turn allows to rephrase  \eqref{extmodE} as
\begin{align}
        \braket{e^{4A}\Tilde{F}-\dd_H (e^{4A-\phi}\nn{Re}\Psi_1),F+\dd_H\Tilde{\ast}_6  (e^{-\phi}\nn{Re}\Psi_1)}&\nonumber\\
        +e^{4A-\phi}\sum_{i \in \nn{loc. sources}}\tau_i\Big[\rho_i^\nn{loc}\nn{vol}_6-\braket{\nn{Re}\Psi_1,j_i}\Big]&=0  \, . 
\label{modEbis}        
        \end{align}
        However, 
for generic SSB configurations the identity 
\eqref{rewritingmeeeq} does not hold and therefore, neither does \eqref{modEbis}. 

For generic SSB configurations without non-calibrated sources, the external component of the Einstein equation therefore can't be reduced further than
\be -\dd(e^{-2\phi}\ast_6\dd e^{4A})=\braket{\Tilde{\ast}_6 \Tilde{F},e^{4A}\Tilde{F}}-\braket{\dd_H\Tilde{\ast}_6 \Tilde{F},e^{4A-\phi}\nn{Re}\Psi_1}. \ee

We can now specify the equations of motion we found above to the SSB class introduced in Subsection \ref{subsec:SSB}, where the pure spinor equations are \eqref{retSSB}, \eqref{calibsourcesssb}, and \eqref{SSBcoord}. 

Let's first discuss further the  external components of the modified Einstein equations. Interestingly, using the gauge BPSness and our specific ansatz for the D-string BPSness violation \eqref{SSBcoord}, we can show that once again the identity \eqref{rewritingmeeeq} holds.
The external modified Einstein equations are therefore again
\begin{align}
        \braket{e^{4A}\Tilde{F}-\dd_H (e^{4A-\phi}\nn{Re}\Psi_1),F+\dd_H\Tilde{\ast}_6  (e^{-\phi}\nn{Re}\Psi_1)}&\nonumber\\
        +e^{4A-\phi}\sum_{i \in \nn{loc. sources}}\tau_i\Big[\rho_i^\nn{loc}\nn{vol}_6-\braket{\nn{Re}\Psi_1,j_i}\Big]&=0,
        \end{align}
        which are satisfied for our class of backgrounds, thanks to the gauge BPSness and the calibration of our space-filling sources.

Moving on to the other equations of motion, the $\{u^{1,2}_{Rm}\}$ terms  in $\Xi$ reduce to\footnote{We could also further specify $\Xi$ by expanding on the SU$(3)\times $SU$(3)$ structure the specific ansatz \eqref{SSBcoord} in the $\Tilde{\ast}_6 \dd_H(e^{2A-\phi}\nn{Im}\Psi_1)$ term, but the resulting expression is neither compact nor enlightening.}
        \begin{align}
            u^1_{Rm}=&(-1)^{|\Psi_2|}e^A(\alpha_m-{\Lambda_m}^n\alpha_n)\\
            u^2_{Rm}=&(-1)^{|\Psi_2|}e^A(\alpha_m-{\Lambda^n}_m\alpha_n).
        \end{align}
    Then, the dilaton equation of motion is automatically satisfied, once the specific form of the D-string BPSness violation \eqref{SSBcoord} is used. The NSNS-field equation of motion doesn't get simplified further, while the internal Einstein equations reduce to
    \be \braket{g_{k(m}\dd y^k\wedge\iota_{n)}\nn{Im}\Psi_1,\dd_H\Xi}=0.\ee
    
We will construct different concrete examples of backgrounds in Section \ref{sec:ex}, and we will see then that the equations of motion presented here can indeed be satisfied by such vacua.

Note also that, as in the one-parameter DWSB case, one can unify the equations of motion for the NSNS-field and the internal Einstein equations, into the following integrated condition. 
     \be \int_{\mathcal{M}}\braket{\delta_{g,B}[\dd_H(e^{2A-\phi}\nn{Im}\Psi_1)],\Xi}=0.\ee
     However, we don't have access to a four-dimensional interpretation of this condition.

 \medskip    

Finally, it is important to realise that for our class of SSB backgrounds,  the effective potential $\mathcal{V}_\text{eff}$
can also be rewritten in terms of the generalised current associated to the generalised foliation. 
Indeed, it is straightforward  to show that\footnote{We prove these equalities using vielbein, since it gives more compact expressions and allows to make the connection with the one-parameter DWSB class in the most natural way, but similar equalities can be derived in the coordinate basis.}
     \begin{align}
          &\braket{\Tilde{\ast}_6 [\dd_H(\text{e}^{2A-\phi}\text{Im}\Psi_1)],\dd_H(\text{e}^{2A-\phi}\text{Im}\Psi_1)}=4\hat{\alpha}_m\hat{\alpha}^m\braket{\Tilde{\ast}_6 j,j}\\
        &(u^1_R)^2=(u^2_R)^2=4\hat{\alpha}_m\hat{\alpha}^me^{-(4A-2\phi)}\frac{|\braket{\Psi_1,j}|^2}{\nn{vol}_6^2}.
     \end{align}
Using these two equations, the effective potential of our  SSB class can be brought to 
           \be  \mathcal{V}_\text{eff}= 2\int_{\mathcal{M}} \hat{\alpha}_m\hat{\alpha}^m\Big[\braket{\Tilde{\ast}_6 j,j}-\frac{|\braket{\Psi_1,j}|^2}{\text{vol}_6}\Big].\ee
           Given that the bound \eqref{boundj} is saturated for our SSB class, the effective potential vanishes again, as expected for Minkowski backgrounds.
           
However, even if this expression is interesting to highlight a common structure between the one-parameter DWSB backgrounds, with effective potential \eqref{veffdwsbj}, and our SSB class, it is also important to remember that the two expressions  come from different terms in the effective potential, and these have drastically different physical interpretations in terms of the effective theory, which we will make more precise later on. 

\medskip

We can now turn to configurations having both SSB and DWSB contributions, with the modified pure spinor equations \eqref{rssbdwsb}, \eqref{alphassbdwsb} and \eqref{RRdwsbssb}. 

The effective potential is 
    \begin{align}
        \mathcal{V}_\text{eff}=
         &\ \frac{1}{2}\int_{\mathcal{M}} e^{-2A}    \braket{\Tilde{\ast}_6 [\dd_H(\text{e}^{3A-\phi}\Psi_2)],\dd_H(\text{e}^{3A-\phi}\bar{\Psi}_2)}\nonumber\\
    &-\frac{1}{4}\int_{\mathcal{M}}e^{-2A}\left(\frac{|\braket{\Psi_1,\dd_H(\text{e}^{3A-\phi}\Psi_2)}|^2}{\text{vol}_6}+\frac{|\braket{\bar{\Psi}_1,\dd_H(\text{e}^{3A-\phi}\Psi_2)}|^2}{\text{vol}_6}\right)\nonumber\\
    &
         +\frac{1}{2}\int_{\mathcal{M}}  \braket{\Tilde{\ast}_6 [\dd_H(\text{e}^{2A-\phi}\text{Im}\Psi_1)],\dd_H(\text{e}^{2A-\phi}\text{Im}\Psi_1)}\nonumber\\
    &-4\int_{\mathcal{M}}\text{vol}_6 e^{4A-2\phi}[(u^1_R)^2+(u^2_R)^2] \, ,
    \label{DWSBSSBpot}
    \end{align}
where the first two lines are the contributions from the violation of the domain-wall calibration condition, while the last two lines correspond to the contributions from the D-string one.

The potential \eqref{DWSBSSBpot} is nothing but the sum of the two effective potentials for the one-parameter DWSB backgrounds and our SSB ansatz from \ref{subsec:SSB}. It can thus be written as
      \be  \mathcal{V}_\text{eff}= \int_{\mathcal{M}} \Big(2\hat{\alpha}_m\hat{\alpha}^m+\frac{1}{2}\text{e}^{-2A}|r|^2\Big)\Big[\braket{\Tilde{\ast}_6 j,j}-\frac{|\braket{\Psi_1,j}|^2}{\text{vol}_6}\Big]\ee
where  the two terms in the bracket still compensate through the saturation of \eqref{boundj}, such that these backgrounds again have vanishing effective potentials.

\medskip

The equations of motion for these backgrounds simply bring together the contributions from the variations of the two effective potentials presented above.

The dilaton equation of motion and the external modified Einstein equation are therefore automatically obeyed given that the modified pure spinor equations are respected. Then the NSNS-field equation of motion is
\be\dd\Big[(-1)^{|\Psi_2|}e^{4A-2\phi}\braket{\nn{Im}(r^\ast\Psi_2),\Theta}_3-e^{2A-\phi}\braket{\nn{Im}\Psi_1,\Xi}_3\Big]=0,\ee
and the internal Einstein equations are
  \be \hspace{-0.1cm}\text{Im}\Big\{(-1)^{|\Psi_2|}e^A\braket{g_{k(m}\dd y^k\wedge\iota_{n)}\Psi_2,\dd_H\big[e^{A-\phi}r^\ast\Theta\big]}\Big\}-\braket{g_{k(m}\dd y^k\wedge\iota_{n)}\nn{Im}\Psi_1,\dd_H\Xi}=0\label{IEEqssbdwsb}.\ee
For compactness, we introduced here the polyform
    \be \Theta=3\nn{Re}\Psi_1+\frac{1}{2}(-1)^{|\Psi_2|}\Lambda_{mn}\gamma^m\nn{Re}\Psi_1\gamma^n \, . 
    \ee

When discussing concrete background examples, the following relationship between the polyforms $\Theta$ and $\Xi$  will prove to be useful
        \begin{align}
            \Xi=&(-1)^{|\Psi_2|}e^{3A-\phi}\hat{\alpha}_a(\gamma^a\Theta+(-1)^{|\Psi_1|+1}\Theta\gamma^a)\\
            =&2(-1)^{|\Psi_2|}e^{3A-\phi}\hat{\alpha}^a\iota_a\Theta.\label{xitheta}
        \end{align}
In contrast with the previous situations, both the NSNS-field equation of motion and the internal Einstein equations could in principle have non-vanishing DWSB and SSB contributions, that could cancel each other out. However, we won't explore this possibility and the background examples that we will present in Section \ref{sec:ex} will have both their DWSB and SSB contributions vanishing independently.

\subsection{Stability and generalised calibrations}\label{subsec:stab}

In the previous sections we showed how to find classes of non-supersymmetric backgrounds by solving modified supersymmetry variations and then considering the additional constraints the configurations must satisfy to be solutions of the equations of motion. 

An important question to address is whether these non-supersymmetric backgrounds are stable. There are two kinds of 
instabilities one could face: under small perturbation or by quantum tunnelling. 

In this paper we try to address the first, namely the potential presence of tachyonic directions. 
As discussed in \cite{Lust:2008zd}, a possible way to answer this question in again looking at the four-dimensional effective 
potentials for the `off-shell' fields of ten-dimensional supergravity around the given configurations. 

Notice that the effective potentials obtained this way are not  genuine potentials associated to four-dimensional theories, since we didn't choose an appropriate truncation for the ten-dimensional modes and we didn't perform the actual reduction.  To do so would require the knowledge of the light modes of the theory, which is complicated to access for general flux vacua\footnote{See for instance \cite{Cassani} for a discussion of the dimensional reduction of general SU$(3)\times $SU$(3)$ type II supergravity backgrounds to $\mathcal{N}=2$ gauged four-dimensional supergravity. It is also worth mentioning here recent work where the full Kaluza-Klein spectrum have been worked out for a variety of flux backgrounds using techniques from Exceptional Field Theory \cite{Malek,Duboeuf}.}. This is beyond the scope of this work. However we will see that their are still some interesting things one can say about our solutions from this approach.

We start by reviewing the analysis of \cite{Lust:2008zd} for 
 the one-parameter DWSB backgrounds and then discuss how to extend it to SBB backgrounds. 

 \medskip

 The idea of \cite{Lust:2008zd} is to go off-shell, which is a way to look at fluctuations around a given solution, 
 and to see whether, under minor constraint on the ten-dimensional supergravity fields, it is possible to derive an effective potential that is positive semi-definite. 

From the general expression   for $\mathcal{V}_\text{eff}$, it is clear that a first constraint to impose is the Bianchi identity \eqref{bianchi} so that the last  line of \eqref{genVeff} vanishes. 
The second condition is to assume that  the parameters 
$\{u^{1,2}_m\}$ in the modified pure spinor equations are zero.\footnote{In term of deformations of the ordinary supersymmetry variations these terms appear in the modified dilatino variations. This condition means that the only allowed deformations of this equation are SU$(3)\times $SU$(3)$ singlets.}  These terms correspond to vector-likes modes of the SU$(3)\times $SU$(3)$ structure group, and should correspond to massive modes from the perspective of the reduced $\mathcal{N}=1$ and   $\mathcal{N}=2$ four-dimensional supergravity theories.

This is particularly clear for  reductions to the  $\mathcal{N}=2$ four-dimensional supergravity. In this case, the vector-like modes correspond to massive spin $\frac{3}{2}$-multiplet degrees of freedom for the four-dimensional theory, see for instance \cite{GranaWaldram}, and these are seen as non-physical degrees of freedom of $\mathcal{N}=2$ four-dimensional supergravity, that should be truncated away.

Anyway, these vector-like modes are not expected to give rise to light or tachyonic contributions to the effective theories.  Therefore by `truncating' them away one is not discarding potential instabilities of the reduced effective theory. 

In order to go off-shell, we still consider an internal geometry that is a generalised foliation, but we do not require the associate generalised submanifold, $(\Pi, \mathcal{R})$,  to be calibrated. 
This means that the violation of the domain-wall calibration condition now takes the more general form
    \be \dd_H(e^{3A-\phi}\Psi_2)=i\Tilde{r}\Tilde{\jmath}_{(\Pi, \mathcal{R})},\ee
    where $\Tilde{\jmath}_{(\Pi, \mathcal{R})}$ is now a generalised current associated to the submanifold $(\Pi, \mathcal{R})$, which isn't necessarily calibrated away from the solution, and 
    $\Tilde{r}$ is just a parameter, eventually identified with the DWSB supersymmetry-breaking parameter mentioned above. 

   The `off-shell' potential is then\footnote{Here we wrote the first line of  \eqref{genVeff} as the square of a polyform, for clarity.}
     \begin{align}
        \mathcal{V}_\text{eff}=&\ 
         \frac{1}{2}\int_{\mathcal{M}}\text{vol}_6  |\dd_H(\text{e}^{2A-\phi}\text{Im}\Psi_1)|^2\nonumber\\
         &+\frac{1}{2}\int_{\mathcal{M}}\text{vol}_6 \ e^{4A}|\Tilde{\ast}_6  F- e^{-4A}\dd_H(e^{4A-\phi}\text{Re}\Psi_1)|^2\nonumber\\
         &+\frac{1}{2}\int_{\mathcal{M}} \text{e}^{-2A}|\Tilde{r}|^2\Big[\braket{\Tilde{\ast}_6 \Tilde{\jmath}_{(\Pi, \mathcal{R})},\Tilde{\jmath}_{(\Pi, \mathcal{R})}}-\frac{|\braket{\Psi_1,\Tilde{\jmath}_{(\Pi, \mathcal{R})}}|^2}{\nn{vol}_6}\Big]\nonumber\\
     &+\sum_{i\subset\text{D-branes}}\tau_i \int_{\mathcal{M}}e^{4A-\phi}(\text{vol}_6 \ \rho_i^\text{loc}-\braket{\text{Re}\Psi_1, j_i}).
    \end{align}
    The terms in the first two lines are trivially positive, and the last two lines are positive because of the calibration bounds \eqref{boundrho} and \eqref{boundj}.
The potential is therefore positive semi-definite, and vanishes precisely for the one-parameter DWSB solutions.

As discussed at length in \cite{Lust:2008zd}, notice that this implies that, under the previous assumptions, the effective potential can naturally be interpreted as being of the no-scale type. This is not  a surprise, given that the one-parameter DWSB class contains the GKP solutions and all its T-duals, which are of the no-scale type \cite{Giddings:2001yu}.

Let us stress  that this is an interesting property of the one-parameter DWSB class, but it is only an argument for the stability of this class with the caveat that we assumed a specific truncation, that we don't have precise control over. 

\medskip

A somewhat similar construction can be found for the backgrounds with SSB discussed in Sections  \ref{subsec:SSB} and  \ref{subsec:ssbdwsb}. Here we discuss
the backgrounds with  SSB and DWSB contributions of \ref{subsec:ssbdwsb}, but the backgrounds of \ref{subsec:SSB} with only SSB contributions exhibit the same behaviour. 

We also impose that the Bianchi identities are satisfied away from the solutions and that  the internal manifold is still a generalised foliation. However we do not truncate away the vector-like modes, as they are fundamental for the SSB constructions. This means that the off-shell violations of the D-string and domain-wall BPSness are
        \begin{align}
  \dd_H(\text{e}^{3A-\phi}\Psi_2)=&\ i\Tilde{r}\Tilde{\jmath}_{(\Pi, \mathcal{R})}\\ 
  \dd_H(\text{e}^{2A-\phi}\text{Im}\Psi_1)=&\ \Tilde{\hat{\alpha}}_m[\hat{\gamma}^m\Tilde{\jmath}_{(\Pi, \mathcal{R})}+(-1)^{|\Tilde{\jmath}_{(\Pi, \mathcal{R})}|}\Tilde{\jmath}_{(\Pi, \mathcal{R})}\hat{\gamma}^m]\label{ssbaway}.
\end{align}
Here $\Tilde{\jmath}_{(\Pi, \mathcal{R})}$ is again a generalised current associated to the submanifold $(\Pi, \mathcal{R})$, which isn't necessarily calibrated away from the solution, and $\{\Tilde{\hat{\alpha}}_m\}$ and $\Tilde{r}$ are just parameters, eventually identified with the SSB and DWSB supersymmetry-breaking parameters.

The `off-shell' potential is then
\begin{align}
        \mathcal{V}_\text{eff}=&\ \frac{1}{2}\int_{\mathcal{M}}\text{vol}_6 \ e^{4A}|\Tilde{\ast}_6  F- e^{-4A}\dd_H(e^{4A-\phi}\text{Re}\Psi_1)|^2\nonumber\\
         &+\int_{\mathcal{M}} \Big(2\Tilde{\hat{\alpha}}_m\Tilde{\hat{\alpha}}^m+\frac{1}{2}\text{e}^{-2A}|\Tilde{r}|^2\Big)\Big[\braket{\Tilde{\ast}_6 \Tilde{\jmath}_{(\Pi, \mathcal{R})},\Tilde{\jmath}_{(\Pi, \mathcal{R})}}-\frac{|\braket{\Psi_1,\Tilde{\jmath}_{(\Pi, \mathcal{R})}}|^2}{\nn{vol}_6}\Big]\nonumber\\
     &+\sum_{i\subset\text{D-branes}}\tau_i \int_{\mathcal{M}}e^{4A-\phi}(\text{vol}_6 \ \rho_i^\text{loc}-\braket{\text{Re}\Psi_1, j_i}) \, .
  \label{offshellVSSB}   
    \end{align}
         The first line is again trivially positive, and the last two lines are positive because of the calibration bounds \eqref{boundrho} and \eqref{boundj}.
The potential \eqref{offshellVSSB} is therefore positive semi-definite, and vanishes precisely for the solutions introduced in \ref{subsec:ssbdwsb}.

However, the situation differs from the one-parameter DWSB case, by the fact that we purposely keep the vector-like modes $\{u^{1,2}_m\}$, that are believed to be massive modes from the effective perspective.

Therefore, we stress that we present this truncation as an interesting property of our backgrounds: there is a `truncation' naturally suggested by the geometry such that the effective potential is positive semi-definite, but this doesn't constitute a proof of perturbative stability, since we have no way to conclude whether we can truncate the ten-dimensional modes in this way or not, and no way to reflect on the relative effective masses between the vector-like modes we kept, and the modes we discarded.

 Moreover, it might not even be sensible to talk about effective theories associated to these ten-dimensional backgrounds, given the presence of these massive vector-like modes, or it could be that their effective theories are non-supersymmetric  with the field content of $\mathcal{N}=1$ or $\mathcal{N}=2$ supergravity with additional massive multiplets, or non-supersymmetric solutions of four-dimensional supergravity theory with higher supersymmetry. Singling out one option among these scenarios would require a rigorous prescription to truncate and reduce the ten-dimensional theory, so we won't address further these questions.
        
\section{Examples of vacua with SSB supersymmetry breaking }\label{sec:ex}

    Up until here we made a rather abstract presentation of our non-supersymmetric backgrounds. The purpose of this section is to present concrete examples of our classes of vacua with SSB contributions. We will focus on internal geometries admitting an SU$(3)$- or a static SU$(2)$-structure, and we will discard the possibility to have a non-trivial two-form $\mathcal{F}$ such that the generalised foliation $(\Sigma,\ \mathcal{F})$ will be entirely defined by the cycle $\Sigma$.

        Throughout this section, we revisit the examples of one-parameter DWSB vacua considered in \cite{Lust:2008zd}, adding SSB contributions to the pure spinor equations and removing or keeping the DWSB one, to construct examples of the class of backgrounds introduced in \ref{subsec:SSB} and \ref{subsec:ssbdwsb} respectively. We will therefore specify the following set of modified pure spinor equations
         \begin{align}
  \dd_H(\text{e}^{3A-\phi}\Psi_2)=&\ 0\ \ \nn{or}\ \ irj\\ 
  \dd_H(\text{e}^{2A-\phi}\text{Im}\Psi_1)=&\ \alpha_m[\gamma^mj+(-1)^{|j|}j\gamma^m]\\
   \dd_H(e^{4A-\phi}\text{Re}\Psi_1)=&\ e^{4A}\Tilde{\ast}_6  F
\end{align}
for our different concrete cases.
        
            We consider the generalised foliations of our internal manifolds to be fibrations
        \be \Sigma\quad\hookrightarrow\quad \mathcal{M}\quad\rightarrow\quad\mathcal{B}\ee
        with $\mathcal{B}$ the base manifold and $\Sigma$ the fibre.

        As discussed in \ref{sec:n0gcg}, the fibres will be calibrated by $\omega^\nn{sf}$ and will be wrapped by the space-filling sources.
\subsection{Type IIB SU$(3)$-structure backgrounds with D5-branes}

The internal manifolds admitting an SU$(3)$-structure have parallel internal spinors $\eta_1$ and $\eta_2$, and in type IIB they have the following pure spinors 
\be \Psi_1=e^{i\theta}e^{iJ}\qquad\Psi_2=e^{-i\theta}\Omega.\ee
We can introduce a local vielbein to write the Kähler form $J$ and the $(3,0)$ form $\Omega$ as
\begin{align}
    J=&-(e^1\wedge e^4+e^2\wedge e^5+e^3\wedge e^6)\\
    \Omega=&(e^1+ie^4)\wedge(e^2+ie^5)\wedge(e^3+ie^6).
\end{align}
Then, following \cite{Martucci:2005ht,Martucci3} one can show that for D5 space-filling branes the algebraic calibration condition \eqref{bound} imposes both that the fibre $\Sigma$ is almost-complex with respect to the almost-complex structure defined by $\Omega$, and 
\be \theta=-\frac{\pi}{2}.\ee
We consider constructions with $2^4$ O5-planes wrapping a two-cycle in the internal geometry, at the fixed point of the $\mathbb{Z}_2$ involutions on the orthogonal four-dimensional space, and we allow for $n_{D5}$ D5-branes, taken to be parallel to these orientifolds.

We will now specialise the discussion to the case of backgrounds with and without DWSB contributions.
\subsubsection{Backgrounds with only SSB contributions}\label{subsubsec:ssbsu3D5}
We can here specify the pure spinor equations \eqref{retSSB}, \eqref{calibsourcesssb} and \eqref{SSBvielbein} satisfied by type IIB SU$(3)$ backgrounds with space-filling D5-branes and with an SSB contribution of the type presented in \ref{subsec:SSB}. They yield
\begin{alignat}{2}
    &H=0\qquad &&e^{2A-\phi}=\nn{const.}\label{ssbsu3IIB}\\
    &F_1=F_5=0\qquad &&\ast_6F_3=-e^{-2\phi}\dd(e^\phi J)\label{rrssbsu3D5}\\
    &\dd(e^A\Omega)=0, && \ \label{retssbd5}
\end{alignat}
with the first line coming from the specific violation of the D-string calibration \eqref{SSB}, and the second and third line coming from gauge and domain-wall BPSness respectively. One condition from \eqref{SSB} is missing here: for now these conditions are the same as the ones one would impose to have a supersymmetric background. If for instance we take the fibres to be along the directions $e^1$ and $e^4$, we have $\nn{vol}_6=\nn{vol}_\Sigma\wedge\nn{vol}_{\mathcal{B}_4}$ with $ \nn{vol}_\Sigma=e^1\wedge e^4$ and $\nn{vol}_{\mathcal{B}_4}=e^2\wedge e^3\wedge e^5\wedge e^6$, and we can write the last condition from \eqref{SSB}, which introduces the breaking of supersymmetry, as
\be e^{2A-\phi}\dd(J\wedge J)=4(\hat{\alpha}_1e^1+\hat{\alpha}_4e^4)\wedge j,\label{ssbviolsu3D5}\ee
with
\be j=4e^{3A-\phi}\nn{vol}_{\mathcal{B}_4}.\ee

Finally, we can specify the equations of motion. The NSNS-field, dilaton and external modified Einstein equations are trivially respected, and to rewrite the internal Einstein equations, it is useful to notice that in the case where $\mathcal{F}=0$, we have
\be \Theta=4(\nn{Re}\Psi_1-\nn{Re}\Psi_1|_\Sigma).\label{simpltheta}\ee 
Using this identity and \eqref{xitheta}, the internal Einstein equations can be shown to reduce to
\be \braket{g_{k(m}\dd y^k\wedge \iota_{n)}J\wedge J,\dd[\hat{\alpha}^a\iota_a(J\wedge J\wedge J)]}=0\qquad a=1,4\ee
which is identically satisfied. We conclude that this family of SSB SU$(3)$ backgrounds with calibrated D5-branes automatically solves its equations of motion, without any further constraints.

   We now turn to the construction of an explicit example of background from this class. We begin by choosing the following metric:
                
                \begin{align}
                \dd s^2=&\text{e}^{2A}\dd s^2_{\mathbb{R}_{1,3}}+  \dd s^2_{\mathcal{M}}\\
                    \dd s^2_{\mathcal{M}}=&\alpha'(2\pi)^2\Big\{\text{e}^{2A}[R_1^2(\eta^1)^2+R_4^2(\eta^4)^2]+\text{e}^{-2A}\sum_{j=2,3,5,6}R^2_j(\dd y^j)^2\Big\},
                \end{align}
               where the warp factor A only depends on the base direction $\mathcal{B}_4 = \{y^2, y^3, y^5, y^6\}$ and $\eta^1$,  $\eta^4$ are non-closed one-forms satisfying

               \be \dd\eta^a=f^a_{ij}\dd y^i\wedge\dd y^j\quad a=1,4\quad i,j=2,3,5,6.\label{vielbeinderiv}\ee
In the constant warp factor limit, this is nothing else then the geometry of a twisted torus. To insure the compactness of $\mathcal{M}$ we take the structure constants $\{f^a_{ij}\}$ to be integer constants, while the radii $\{R_a,R_i\}$ can take any real value.
               The Kähler form $J$ and the $(3,0)$ form $\Omega$ are
               \begin{align}
                   J=&-\alpha'(2\pi)^2[e^{2A}R_1R_4\eta^1\wedge\eta^4+e^{-2A}(R_2R_5\dd y^2\wedge\dd y^5+R_3R_6\dd y^3\wedge\dd y^6)]\\
                   \Omega=&{\alpha'}^{3/2}(2\pi)^3 e^{-A}(R_1\eta^1+i R_4\eta^4)\wedge(R_2\dd y^2+i R_5\dd y^5)\wedge(R_3\dd y^3+i R_6\dd y^6).
               \end{align}
               
The domain-wall BPSness \eqref{retssbd5} now takes the form
 \begin{align}
            f^1_{26} R_1 R_3 R_5 -f^4_{56} R_2 R_3 R_4  - f^1_{35} R_1 R_2 R_6 + f^4_{23 }R_4 R_5 R_6=&0\label{noDWSB1}\\
            f^4_{26} R_3 R_4 R_5+f^1_{56} R_1 R_2 R_3  - f^4_{35} R_2 R_4 R_6 - f^1_{23} R_1 R_5 R_6=&0\label{noDWSB2},
        \end{align}   
        and the RR-fluxes \eqref{rrssbsu3D5} read
        \begin{align}
        F_3=&e^{2A-\phi}[\ast_{\mathcal{B}_4}\dd e^{-4A}-\alpha'(2\pi)^2\left(R_1^2\eta^1\wedge\ast_\mathcal{
        B}\dd\eta^1+R_4^2\eta^4\wedge\ast_\mathcal{
        B}\dd\eta^4\right)],\label{rrfluxssbsu3D5}\\
            \equiv&e^{2A-\phi}\ast_{\mathcal{B}_4}\dd e^{-4A}+\alpha'(2\pi)^2F_3^\nn{bg}
        \end{align} 
        with $\ast_{\mathcal{B}_4}$ the four-dimensional Hodge operator on the unwarped base. 
        The first term comes from the back-reaction of the D5-branes and O5-planes, while the second term should be thought as a properly quantised background flux, so it must be an integer valued three-form, which constrains the radii $R_i$ and the value of $e^{2A-\phi}$.

            The generalised current associated to the sources is
            \be j=4\alpha'^2 (2\pi)^4e^{-A-\phi} R_2 R_3 R_5 R_6  \dd y^2\wedge \dd y^3\wedge \dd y^5\wedge \dd y^6,\ee
            and the D-string BPSness violation \eqref{ssbviolsu3D5} is
            \be  e^{2A-\phi}\dd(J\wedge J)=8\pi\sqrt{\alpha'}e^A(R_1\hat{\alpha}_1\eta^1+R_4\hat{\alpha}_4\eta^4)\wedge j,\label{jsu3D5}\ee
        which is satisfied for the following supersymmetry breaking parameters
        \begin{align}
            \hat{\alpha}_1=&\ e^{2A}\frac{R_4}{16\pi\sqrt{\alpha'}}\Big(\frac{f^4_{36}}{R_3R_6}+\frac{f^4_{25}}{R_2R_5}\Big)\\
        \hat{\alpha}_4=&-e^{2A}\frac{R_1 }{16\pi\sqrt{\alpha'}}\Big(\frac{f^1_{36}}{R_3R_6}+\frac{f^1_{25}}{R_2R_5}\Big).
        \end{align}
        Finally, we can specify the Bianchi identities \eqref{bianchi}, which in this case are only non trivial for $F_3$\footnote{Note that the current \eqref{jsu3D5} and the one in \eqref{currentdef} differ by an overall factor.}
         \begin{align} 
         \dd F_3&=e^{2A-\phi}\big[-\nabla^2_{\mathcal{B}_4}e^{-4A}+\frac{Y}{4\alpha'(2\pi)^2}\big]\text{v}\Tilde{\text{o}}\text{l}_{\mathcal{B}_4}\\
         &=\frac{e^{2A-\phi}}{\alpha'(2\pi)^2}\Big(\sum_{i=1}^{16}\delta^4_{\mathcal{B}_4}(y_i)-\sum_{j=1}^{n_{D5}}\delta^4_{\mathcal{B}_4}(y_j)\Big)\text{v}\Tilde{\text{o}}\text{l}_{\mathcal{B}_4},
         \end{align}
       where $\text{v}\Tilde{\text{o}}\text{l}_{\mathcal{B}_4}$ is the unwarped volume form on $\mathcal{B}_4$, the charges of the O5 and D5 sources have been normalised to $-1$ and $1$ respectively, and with
       \begin{align}
         Y&=\frac{(f^1_{23})^2}{(R_2R_3)^2}+\frac{(f^1_{25})^2}{(R_2R_5)^2}+\frac{(f^1_{26})^2}{(R_2R_6)^2}+\frac{(f^1_{35})^2}{(R_3R_5)^2}+\frac{(f^1_{36})^2}{(R_3R_6)^2}+\frac{(f^1_{56})^2}{(R_5R_6)^2}\nonumber\\
         &\quad+\frac{(f^4_{23})^2}{(R_2R_3)^2}+\frac{(f^4_{25})^2}{(R_2R_5)^2}+\frac{(f^4_{26})^2}{(R_2R_6)^2}+\frac{(f^4_{35})^2}{(R_3R_5)^2}+\frac{(f^4_{36})^2}{(R_3R_6)^2}+\frac{(f^4_{56})^2}{(R_5R_6)^2}.
         \end{align}
         The corresponding tadpole condition connects the sources to the radii and structure constants:
         \be n_{D5}+\frac{Y}{4}=16.\ee
      \subsubsection{Backgrounds with both SSB and DWSB contributions} 
       
       One can construct a similar class of SU$(3)$ backgrounds with space-filling D5-branes, with an additional supersymmetry breaking contribution from the violation of the domain-wall BPSness. Considering again the fibres to be along the directions $e^1$ and $e^4$, the pure spinor equations \eqref{rssbdwsb}, \eqref{alphassbdwsb} and \eqref{RRdwsbssb} take the form
       \begin{alignat}{2}
    &H=0\qquad &&e^{2A-\phi}=\nn{const.}\label{ssbdwsbsu3IIB}\\
    &F_1=F_5=0\qquad &&\ast_6F_3=-e^{-2\phi}\dd(e^\phi J)\label{rrssbdwsbsu3D5}
\end{alignat}

and \begin{align}
    e^{2A-\phi}\dd(J\wedge J)&=4(\hat{\alpha}_1e^1+\hat{\alpha}_4e^4)\wedge j\label{dstringviolssbdwsbsu3D5}\\
     e^{2A-\phi}\dd(e^A\Omega)&=irj\label{dwviolssbdwsbsu3D5}
\end{align}
with again
\be j=4e^{3A-\phi}\nn{vol}_{\mathcal{B}_4}.\ee
Turning to the equations of motion, the NSNS-field equation is still trivially satisfied, such as the dilaton equation, and the external modified Einstein equations. Therefore only the internal Einstein equation has an additional non-trivial contribution. From \eqref{simpltheta}, we see that it reduces to
\be \nn{Re}\braket{g_{k(m}\dd y^k\wedge \iota_{n)}\Omega,\dd(e^{-A}r^\ast J|_{\mathcal{B}_4})}=0.\label{Einsteineomssbdwsb}\ee
Since $g_{k(m}\dd y^k\wedge \iota_{n)}\Omega$ is either a $(3, 0)$ or a primitive $(2, 1)$ form, imposing 
\be \left[\dd(e^{-A}rJ|_{\mathcal{B}_4})\right]^{3,0}= \left[\dd(e^{-A}rJ|_{\mathcal{B}_4})\right]^{2,1}_\nn{prim}=0\label{condinteistein}\ee
is enough to satisfy \eqref{Einsteineomssbdwsb}. If we consider $r$ and the warp factor to be constant along the fibre, as is usual when the localised sources wrap the fibres, and if we have $\dd J|_{\mathcal{B}_4}=f\wedge J|_{\mathcal{B}_4}$ with $f$ a real function on the base, then the conditions \eqref{condinteistein} amounts to
\be \left[
\dd(e^{-A}rf)\right]^{1,0}=0.\ee

Let us construct an example of these backgrounds, which will respect this condition. 

We start off with the same metric ansatz as for the case with only the SSB contribution
 \begin{align}
                \dd s^2=&\text{e}^{2A}\dd s^2_{\mathbb{R}_{1,3}}+  \dd s^2_{\mathcal{M}}\\
                    \dd s^2_{\mathcal{M}}=&\alpha'(2\pi)^2\Big\{\text{e}^{2A}[R_1^2(\eta^1)^2+R_4^2(\eta^4)^2]+\text{e}^{-2A}\sum_{j=2,3,5,6}R^2_j(\dd y^j)^2\Big\},
                \end{align}      
         with the one forms $\eta^{1}$ and $\eta^4$ respecting again \eqref{vielbeinderiv}. The Kähler form $J$ and the $(3,0)$ form $\Omega$ are once more
          \begin{align}
                   J=&-\alpha'(2\pi)^2[e^{2A}R_1R_4\eta^1\wedge\eta^4+e^{-2A}(R_2R_5\dd y^2\wedge\dd y^5+R_3R_6\dd y^3\wedge\dd y^6)]\\
                   \Omega=&{\alpha'}^{3/2}(2\pi)^3 e^{-A}(R_1\eta^1+i R_4\eta^4)\wedge(R_2\dd y^2+i R_5\dd y^5)\wedge(R_3\dd y^3+i R_6\dd y^6).
            \end{align}
    The gauge BPSness \eqref{rrssbdwsbsu3D5} yields the following RR-flux
       \begin{align}
        F_3=&e^{2A-\phi}[\ast_{\mathcal{B}_4}\dd e^{-4A}-\alpha'(2\pi)^2\left(R_1^2\eta^1\wedge\ast_\mathcal{
        B}\dd\eta^1+R_4^2\eta^4\wedge\ast_\mathcal{
        B}\dd\eta^4\right)],\\
            \equiv&e^{2A-\phi}\ast_{\mathcal{B}_4}\dd e^{-4A}+\alpha'(2\pi)^2F_3^\nn{bg}
        \end{align} 
    Even though the form of the RR-flux is similar to the background with only an SSB contribution \eqref{rrfluxssbsu3D5}, we stress that their components differ, since the structure constants $\{f^a_{ij}\}$ are different. Indeed, here we don't respect the conditions \eqref{noDWSB1} and \eqref{noDWSB2} since we have a DWSB contribution. Besides, the values of $e^{2A-\phi}$ and the radii $R_i$ should be chosen appropriately such that the background fluxes are integer-valued.

        The generalised current is again
            \be j=4\alpha'^2 (2\pi)^4e^{-A-\phi} R_2 R_3 R_5 R_6  \dd y^2\wedge \dd y^3\wedge \dd y^5\wedge \dd y^6,\ee
            and the D-string \eqref{dstringviolssbdwsbsu3D5} and domain-wall \eqref{dwviolssbdwsbsu3D5} BPSness violations are
            \begin{align}
      e^{2A-\phi}\dd(J\wedge J)=&8\pi\sqrt{\alpha'}e^A(R_1\hat{\alpha}_1\eta^1+R_4\hat{\alpha}_4\eta^4)\wedge j\\
     e^{2A-\phi}\dd(e^A\Omega)=&irj
\end{align}
which are satisfied if
\begin{align}
            r=&\ \frac{e^{3A}}{8\pi\sqrt{\alpha'}}\Big(\frac{R_1f^1_{56}+i R_4f^4_{56}}{R_5R_6}-\frac{R_1f^1_{23}+i R_4f^4_{23}}{R_2R_3}\\
            &+\frac{i R_1f^1_{35}- R_4f^4_{35}}{R_3R_5}-\frac{i R_1f^1_{26}- R_4f^4_{26}}{R_2R_6}\Big)\\
               \hat{\alpha}_1=&\ e^{2A}\frac{R_4}{16\pi\sqrt{\alpha'}}\Big(\frac{f^4_{36}}{R_3R_6}+\frac{f^4_{25}}{R_2R_5}\Big)\\
        \hat{\alpha}_4=&-e^{2A}\frac{R_1 }{16\pi\sqrt{\alpha'}}\Big(\frac{f^1_{36}}{R_3R_6}+\frac{f^1_{25}}{R_2R_5}\Big).
        \end{align}
        Turning to the equations of motion, we have
        \be \dd(e^{-A}rJ|_{\mathcal{B}_4})=0\ee 
        so \eqref{condinteistein} are obeyed and the internal Einstein equations are satisfied. As discussed above, every other equations of motion are satisfied.

            Finally, we can write down the only non-trivial Bianchi identity, the one for the RR-flux
            \begin{align} 
         \dd F_3&=e^{2A-\phi}\big[-\nabla^2_{\mathcal{B}_4}e^{-4A}+\frac{Y}{4\alpha'(2\pi)^2}\big]\text{v}\Tilde{\text{o}}\text{l}_{\mathcal{B}_4}\\
         &=\frac{e^{2A-\phi}}{\alpha'(2\pi)^2}\Big(\sum_{i=1}^{16}\delta^4_{\mathcal{B}_4}(y_i)-\sum_{j=1}^{n_{D5}}\delta^4_{\mathcal{B}_4}(y_j)\Big)\text{v}\Tilde{\text{o}}\text{l}_{\mathcal{B}_4},
         \end{align}
    with
       \begin{align}
         Y&=\frac{(f^1_{23})^2}{(R_2R_3)^2}+\frac{(f^1_{25})^2}{(R_2R_5)^2}+\frac{(f^1_{26})^2}{(R_2R_6)^2}+\frac{(f^1_{35})^2}{(R_3R_5)^2}+\frac{(f^1_{36})^2}{(R_3R_6)^2}+\frac{(f^1_{56})^2}{(R_5R_6)^2}\nonumber\\
         &\quad+\frac{(f^4_{23})^2}{(R_2R_3)^2}+\frac{(f^4_{25})^2}{(R_2R_5)^2}+\frac{(f^4_{26})^2}{(R_2R_6)^2}+\frac{(f^4_{35})^2}{(R_3R_5)^2}+\frac{(f^4_{36})^2}{(R_3R_6)^2}+\frac{(f^4_{56})^2}{(R_5R_6)^2}.
         \end{align}
        Once again, this expression is similar to the one for the background with only SSB contribution presented in \ref{subsubsec:ssbsu3D5}, but the addition of the DWSB contribution actually modifies the structure constants $\{f^a_{ij}\}$ upon releasing the constraints \eqref{noDWSB1} and \eqref{noDWSB2}, which also alters the corresponding tadpole condition
         \be n_{D5}+\frac{Y}{4}=16.\ee
\subsection{Type IIA SU$(3)$-structure backgrounds with D6-branes}
Type IIA SU$(3)$ backgrounds have the following pure spinors
\be \Psi_1=\Omega\qquad\Psi_2=e^{-i\theta}e^{iJ}.\ee
Then, following \cite{Martucci:2005ht,Martucci3}, one can show that the algebraic calibration condition \eqref{bound} for D6 space-filling branes wrapping an internal cycle $\Sigma$ imposes
\be J|_\Sigma=0\qquad\nn{Im}\Omega|_\Sigma=0.\ee
We consider constructions with $2^3$ O6-planes wrapping a two-cycle in the internal geometry, at the fixed point of the $\mathbb{Z}_2$ involutions on the orthogonal three-dimensional space, and we allow for $n_{D6}$ D6-branes, taken to be parallel to these orientifolds.

We again introduce a local vielbein and express the Kähler and $(3,0)$ form as
\begin{align}
    J=&-(e^1\wedge e^4+e^2\wedge e^5+e^3\wedge e^6)\\
    \Omega=&(e^1+ie^4)\wedge(e^2+ie^5)\wedge(e^3+ie^6),
\end{align}
and we consider the fibres wrapped by the sources to be along the $e^1$, $ e^2$ and $e^3$ directions. We then have $\nn{vol}_6=\nn{vol}_\Sigma\wedge\nn{vol}_{\mathcal{B}_3}$ with $ \nn{vol}_\Sigma=e^1\wedge e^2\wedge e^3$ and $\nn{vol}_{\mathcal{B}_3}=e^4\wedge e^5\wedge e^6$. We can now write the pure spinor equations obeyed by these backgrounds, for the cases with and without a DWSB contribution.
\subsubsection{Backgrounds with only SSB contributions}\label{subsubsec:ssbsu3D6}
The pure spinor equations \eqref{retSSB}, \eqref{calibsourcesssb} and \eqref{SSBvielbein}, which correspond to having only an SSB contribution, read
\begin{alignat}{2}
     &e^{3A-\phi}=\nn{const.}\qquad &&e^{i\theta}=\nn{const.}\\
     &H=0\qquad&&\dd J=0\label{dwbpsssbiiA}
\end{alignat}
for the domain-wall BPSness,
\begin{align}
    F_0=&\ F_4=\ F_6\\
    \ast F_2=&-e^{-4A}\dd(e^{4A-\phi}\nn{Re}\Omega)\label{fluxesssbIIA}
\end{align}
for the gauge BPSness, and
\be e^{3A-\phi}\dd(e^{-A}\nn{Im}\Omega)=2(\hat{\alpha}_1e^1+\hat{\alpha}_2e^2+\hat{\alpha}_3e^3)\wedge j\label{ssbIIA}\ee
with
\be j=-4e^{3A-\phi}\nn{vol}_{\mathcal{B}_3}\ee
for the violation of the D-string BPSness.

    Turning to the equations of motion, the internal Einstein equations are
\be \braket{g_{k(m}\dd y^k\wedge \iota_{n)}\nn{Im}\Omega,\dd\big[\hat{\alpha}^a\iota_a( \nn{Re}\Omega-\nn{Re}\Omega|_\Sigma)\big]}=0\qquad a=1,2,3.\label{EEQSSBIIA}\ee
It is natural to construct fibered backgrounds respecting
\be \dd\big[\hat{\alpha}^a\iota_a( \nn{Re}\Omega-\nn{Re}\Omega|_\Sigma)\big]=0\qquad a=1,2,3\ee
and thus satisfying the internal Einstein equations. We will now construct such a background.

    We start with the following metric ansatz
      \begin{align}
                \dd s^2=&\text{e}^{2A}\dd s^2_{\mathbb{R}_{1,3}}+  \dd s^2_{\mathcal{M}}\\
                    \dd s^2_{\mathcal{M}}=&\alpha'(2\pi)^2\Big\{\text{e}^{2A}\sum_{a=1,2,3}R_a^2(\eta^a)^2+\text{e}^{-2A}\sum_{j=4,5,6}R^2_j(\dd y^j)^2\Big\}
                \end{align}
 where as in the type IIB case, the warp factor A only depends on the base direction $\mathcal{B}_3 = \{ y^4, y^5, y^6\}$ and $\eta^1$, $\eta^2$ and $\eta^2$ are non-closed one-forms satisfying

               \be \dd\eta^a=f^a_{ij}\dd y^i\wedge\dd y^j\quad a=1,2,3\quad i,j=4,5,6.\label{vielbeinderiv2}\ee
In the constant warp factor limit, this again corresponds to the geometry of a twisted torus. To insure the compactness of $\mathcal{M}$, the structure constants $\{f^a_{ij}\}$ have to be integer constants, and the radii $\{R_a,R_i\}$ can take any real value.
               The Kähler form $J$ and the $(3,0)$ form $\Omega$ are
                
 \begin{alignat}{2}
    J=&-\alpha'(2\pi)^2&&[R_1R_4\eta^1\wedge\dd y^4+R_2R_5\eta^2\wedge\dd y^5+R_3R_6\eta^3\wedge\dd y^6]\\
                   \Omega=&{\alpha'}^{3/2}(2\pi)^3&&(e^{A}R_1\eta^1+ie^{-A}R_4\dd y^4)\wedge(e^{A}R_2\eta^2+i e^{-A}R_5\dd y^5)\\
                &\   &&\wedge(e^{A}R_3\eta^3+ie^{-A} R_6\dd y^6).
            \end{alignat}    

        The domain-wall BPSness \eqref{dwbpsssbiiA} reduces to 
        \be R_1R_4f^1_{56}+ R_2R_5f^2_{64}+ R_3R_6f^3_{45}=0,\label{DWBPSnesssu3D6}\ee
and the RR-fluxes \eqref{fluxesssbIIA} read
\begin{align}
     F_2=&e^{3A-\phi}\big[\ast_{\mathcal{B}_3}\dd e^{-4A}-\alpha'(2\pi)^2\sum_{a=1,2,3}R^2_a\eta^a\wedge\ast_{\mathcal{B}_3}\dd\eta^a\big]\\
     F_2\equiv&e^{3A-\phi}\ast_{\mathcal{B}_3}\dd e^{-4A}+\alpha'(2\pi)^2 F_2^\nn{bg},
\end{align}
with $\ast_{\mathcal{B}_3}$ the three-dimensional Hodge operator on the unwarped base. As in the type IIB case, the background RR-fluxes must be integer-valued, constraining the values of the radii $R_i$ and of $e^{3A-\phi}$.

    The generalised current is
    \be j=-4\alpha'^{3/2}(2\pi)^3e^{-\phi}R_4R_5R_6\dd y^4\wedge \dd y^5\wedge \dd y^6,\ee
and the D-string BPSness violation \eqref{ssbIIA} is
\be e^{3A-\phi}\dd(e^{-A}\nn{Im}\Omega)=4\pi\sqrt{\alpha'}e^A(R_1\hat{\alpha}_1\eta^1+R_2\hat{\alpha}_2\eta^2+R_3\hat{\alpha}_3\eta^3)\wedge j\ee
which gives
\begin{align}
    \hat{\alpha}_1=&\ \frac{e^{2A}}{16\pi\sqrt{\alpha'}R_4}\left(\frac{R_3}{R_6}f^3_{46}+\frac{R_2}{R_5}f^2_{45}\right)\\
    \hat{\alpha}_2=&\ \frac{e^{2A}}{16\pi\sqrt{\alpha'}R_5}\left(\frac{R_3}{R_6}f^3_{56}-\frac{R_1}{R_4}f^1_{45}\right)\\
    \hat{\alpha}_3=&-\frac{e^{2A}}{16\pi\sqrt{\alpha'}R_6}\left(\frac{R_2}{R_5}f^2_{56}+\frac{R_1}{R_4}f^1_{46}\right).
\end{align}

With these $\hat{\alpha}$'s, one can show that
\be \dd\big[\hat{\alpha}^a\iota_a( \nn{Re}\Omega-\nn{Re}\Omega|_\Sigma)\big]=0\qquad a=1,2,3,\ee
and the internal Einstein equations \eqref{EEQSSBIIA} are therefore satisfied.

    Finally, the Bianchi identities reduce to
  \begin{align}
      \dd F_2&=e^{3A-\phi}\big[-\nabla^2_{\mathcal{B}_3}e^{-4A}+\frac{Z}{2\alpha'(2\pi)^2}\big]\text{v}\Tilde{\text{o}}\text{l}_{\mathcal{B}_3}\nonumber\\
      &=\frac{e^{3A-\phi}}{\alpha'(2\pi)^2}\Big(\sum_{i=1}^{8}\delta^4_{\mathcal{B}_4}(y_i)-\sum_{j=1}^{n_{D6}}\delta^4_{\mathcal{B}_4}(y_j)\Big)\text{v}\Tilde{\text{o}}\text{l}_{\mathcal{B}_4},
      \end{align}
  with $\text{v}\Tilde{\text{o}}\text{l}_{\mathcal{B}_3}$ the unwarped volume form on $\mathcal{B}_3$, the charges of the O6 and D6 sources have been normalised to $-1$ and $1$ respectively, and with
  \begin{align} Z&=\frac{(f^1_{45})^2}{(R_4R_5)^2}+\frac{(f^1_{46})^2}{(R_4R_6)^2}+\frac{(f^1_{56})^2}{(R_5R_6)^2}+\frac{(f^2_{45})^2}{(R_4R_5)^2}+\frac{(f^2_{46})^2}{(R_4R_6)^2}+\frac{(f^2_{56})^2}{(R_5R_6)^2}\nonumber\\
  &\quad+\frac{(f^3_{45})^2}{(R_4R_5)^2}+\frac{(f^3_{46})^2}{(R_4R_6)^2}+\frac{(f^3_{56})^2}{(R_5R_6)^2}.
  \end{align}
   The corresponding tadpole condition connects the sources to the radii and structure constants:
         \be n_{D6}+\frac{Z}{2}=8.\ee
    \subsubsection{Backgrounds with both SSB and DWSB contributions} 
    One can construct a similar class of 
    SU$(3)$ backgrounds that has space-filling D6-branes,
with an additional supersymmetry breaking contribution from the violation of the domain-
wall BPSness. In this case, the pure
spinor equations \eqref{rssbdwsb}, \eqref{alphassbdwsb} and \eqref{RRdwsbssb} take the form
\begin{alignat}{2}
    &e^{3A-\phi}=\nn{const.}\qquad e^{i\theta}=\nn{const.}\qquad&&F_4=F_6=0\\
    &\ast F_2=-e^{-4A}\dd(e^{4A-\phi}\nn{Re}\Omega)\qquad&& \ast F_0=e^{-\phi}H\wedge\nn{Re}\Omega\label{fluxesssbdwsbIIA}
\end{alignat}
and \begin{align}
    &e^{3A-\phi}e^{-i\theta}(H+i\dd J)=irj\\
    &e^{3A-\phi}\dd(e^{-A}\nn{Im}\Omega)=2(\hat{\alpha}_1e^1+\hat{\alpha}_2e^2+\hat{\alpha}_3e^3)\wedge j
\end{align}
with again
\be j=-4e^{3A-\phi}\nn{vol}_{\mathcal{B}_3}.\ee
 Turning to the equations of motion, the internal Einstein equations are
   \be \text{Re}\Big\{e^Ae^{-i\theta}\braket{g_{k(m}\dd y^k\wedge\iota_{n)}J,\dd_H\big[e^{-2A}r^\ast\Theta\big]}\Big\}+\braket{g_{k(m}\dd y^k\wedge\iota_{n)}\nn{Im}\Omega,\dd_H(\hat{\alpha}^a\iota_a\Theta)}=0,\ee
with $a=1,2,3$ and again
\be \Theta=4( \nn{Re}\Omega-\nn{Re}\Omega|_\Sigma).\ee

It is sufficient to obey
\begin{align}
    \dd_H\big[e^{-2A}r^\ast\Theta\big]=&\ 0\label{eeq1}\\
    \dd_H(\hat{\alpha}^a\iota_a\Theta)=&\ 0\qquad a=1,2,3\label{eeq2}
\end{align}
to satisfy the internal Einstein equations, and we will shortly turn to the construction of an example background respecting \eqref{eeq1} and \eqref{eeq2}.
As for the NSNS-field equation of motion, it reduces exactly to \eqref{eeq1}.

    We start the construction of an explicit background belonging to this class with considering the following NSNS ansatz
     \begin{align}
                \dd s^2=&\text{e}^{2A}\dd s^2_{\mathbb{R}_{1,3}}+  \dd s^2_{\mathcal{M}}\\
                    \dd s^2_{\mathcal{M}}=&\alpha'(2\pi)^2\Big\{\text{e}^{2A}\sum_{a=1,2,3}R_a^2(\eta^a)^2+\text{e}^{-2A}\sum_{j=4,5,6}R^2_j(\dd y^j)^2\Big\}\\
                    H=&\alpha'(2\pi)^2\Big(N \dd y^4\wedge\dd y^5\wedge\dd y^6+\sum_{a=1,2,3}B_a\dd\eta^a\wedge\dd y^{a+3}\Big)
                \end{align}
 where $N\in\mathbb{Z}$, $B_a\in\mathbb{Z}$ and again the warp factor A only depends on the base direction $\mathcal{B}_3 = \{ y^4, y^5, y^6\}$ and $\eta^1$, $\eta^2$ and $\eta^2$ are non-closed one-forms satisfying \eqref{vielbeinderiv2}. The Kähler form $J$ and the $(3,0)$ form $\Omega$ are again
                
 \begin{alignat}{2}
    J=&-\alpha'(2\pi)^2&&[R_1R_4\eta^1\wedge\dd y^4+R_2R_5\eta^2\wedge\dd y^5+R_3R_6\eta^3\wedge\dd y^6]\\
                   \Omega=&{\alpha'}^{3/2}(2\pi)^3&&(e^{A}R_1\eta^1+ie^{-A}R_4\dd y^4)\wedge(e^{A}R_2\eta^2+i e^{-A}R_5\dd y^5)\\
                &\   &&\wedge(e^{A}R_3\eta^3+ie^{-A} R_6\dd y^6).
            \end{alignat}    
The RR-fluxes \eqref{fluxesssbdwsbIIA} read
\begin{align} 
F_0=&-e^{3A-\phi}\frac{R_1R_2R_3}{2\pi\sqrt{\alpha'}\nn{Vol}(\mathcal{M})}\big[N+B_1f^1_{56}-B_2f^2_{46}+B_3f^3_{45}\big]\\
F_2=&\ e^{3A-\phi}\big[\ast_{\mathcal{B}_3}\dd e^{-4A}-\alpha'(2\pi)^2\sum_{a=1,2,3}R^2_a\eta^a\wedge\ast_{\mathcal{B}_3}\dd\eta^a\big]\\
F_2\equiv&e^{3A-\phi}\ast_{\mathcal{B}_3}\dd e^{-4A}+\alpha'(2\pi)^2 F_2^\nn{bg},
,\end{align}
with the internal manifold volume $\nn{Vol}(\mathcal{M})$ normalised in $\alpha'$ units. Both $F_0$ and $F_2^\nn{bg}$ are background fluxes and must be integer-valued, constraining the radii and $ e^{3A-\phi}$ again.
As before, the generalised current is
    \be j=-4\alpha'^{3/2}(2\pi)^3e^{-\phi}R_4R_5R_6\dd y^4\wedge \dd y^5\wedge \dd y^6,\ee
and the domain-wall and D-string BPSness violations are
\begin{align}
    &e^{3A-\phi}e^{-i\theta}(H+i\dd J)=irj\\
    &e^{3A-\phi}\dd(e^{-A}\nn{Im}\Omega)=4\pi\sqrt{\alpha'}e^A(R_1\hat{\alpha}_1\eta^1+R_2\hat{\alpha}_2\eta^2+R_3\hat{\alpha}_3\eta^3)\wedge j
\end{align}
and are satisfied if
\begin{align}
    r=\ &\frac{e^{3A}e^{-i\theta}}{8\pi\sqrt{\alpha'}R_4R_5R_6}\Big[i\big(N+B_1f^1_{56}-B_2f^2_{46}+B_3f^3_{45}\big)\\
    &+f^1_{56}R_1R_4-f2_{46}R_2R_5+f3_{45}R_3R_6\Big]\\
    \hat{\alpha}_1=&\ \frac{e^{2A}}{16\pi\sqrt{\alpha'}R_4}\left(\frac{R_3}{R_6}f^3_{46}+\frac{R_2}{R_5}f^2_{45}\right)\\
    \hat{\alpha}_2=&\ \frac{e^{2A}}{16\pi\sqrt{\alpha'}R_5}\left(\frac{R_3}{R_6}f^3_{56}-\frac{R_1}{R_4}f^1_{45}\right)\\
    \hat{\alpha}_3=&-\frac{e^{2A}}{16\pi\sqrt{\alpha'}R_6}\left(\frac{R_2}{R_5}f^2_{56}+\frac{R_1}{R_4}f^1_{46}\right).
\end{align}
With these supersymmetry breaking parameters, one can show that \eqref{eeq1} and \eqref{eeq2} are satisfied, and therefore the NSNS-field and internal metric equations of motion are obeyed.
    
        Finally, the Bianchi identities for the RR-fluxes read
  \begin{align}
      \dd F_2&=e^{3A-\phi}\big[-\nabla^2_{\mathcal{B}_3}e^{-4A}+\frac{Z}{2\alpha'(2\pi)^2}\big]\text{v}\Tilde{\text{o}}\text{l}_{\mathcal{B}_3}\nonumber\\
      &=\frac{e^{3A-\phi}}{\alpha'(2\pi)^2}\Big(\sum_{i=1}^{8}\delta^4_{\mathcal{B}_4}(y_i)-\sum_{j=1}^{n_{D6}}\delta^4_{\mathcal{B}_4}(y_j)\Big)\text{v}\Tilde{\text{o}}\text{l}_{\mathcal{B}_4},
      \end{align}
with again
  \begin{align} Z&=\frac{(f^1_{45})^2}{(R_4R_5)^2}+\frac{(f^1_{46})^2}{(R_4R_6)^2}+\frac{(f^1_{56})^2}{(R_5R_6)^2}+\frac{(f^2_{45})^2}{(R_4R_5)^2}+\frac{(f^2_{46})^2}{(R_4R_6)^2}+\frac{(f^2_{56})^2}{(R_5R_6)^2}\nonumber\\
  &\quad+\frac{(f^3_{45})^2}{(R_4R_5)^2}+\frac{(f^3_{46})^2}{(R_4R_6)^2}+\frac{(f^3_{56})^2}{(R_5R_6)^2}.
  \end{align}
  This expression is similar to the one for the background with only SSB contribution presented in \ref{subsubsec:ssbsu3D6}, but the addition of the DWSB contribution actually modifies the structure constants $\{f^a_{ij}\}$ upon releasing the constraint \eqref{DWBPSnesssu3D6}, which also alters the corresponding tadpole condition
         \be n_{D6}+\frac{Z}{2}=8.\ee
\subsection{Type IIB SU$(2)$-structure backgrounds with D5-branes}

We now turn to the discussion of type IIB backgrounds admitting a static SU$(2)$ structure and having space-filling D5-branes. The two internal spinors $\eta_1$ and $\eta_2$ of such backgrounds are everywhere orthogonal, which means that one can specify a one-form $\theta=\theta_m\dd y^m$ such that
\be \eta_2=-\frac{i}{2}\theta_m\gamma^m\eta_1^\ast.\ee

It is natural to parametrise the two SU$(3)$ structures defined by $\eta_1$ and $\eta_2$ as
\begin{alignat}{2}
    J_1=&-\frac{i}{2}\theta\wedge\bar{\theta}+\mathfrak{j}\qquad  \Omega_1=&&-\theta\wedge w\\
    J_2=&-\frac{i}{2}\theta\wedge\bar{\theta}-\mathfrak{j}\qquad  \Omega_2=&&\ \theta\wedge\bar{w}
\end{alignat}
with $\iota_\theta\mathfrak{j}=\iota_{\bar{\theta}}\mathfrak{j}=\iota_\theta w=\iota_{\bar{\theta}}w=0$. The corresponding pure spinors are
\begin{align}
    \Psi_1=&\ w \wedge e^{\frac{1}{2}\theta\wedge\bar{\theta}}\\
    \Psi_2=&\ \theta\wedge e^{i\mathfrak{j}}.
\end{align}
Following \cite{Martucci3}, one can show that the calibration of the space-filling D5-branes wrapping a cycle $\Sigma$ imposes
\be \theta|_\Sigma=0\qquad \mathfrak{j}|_\Sigma\qquad\mathcal{F}=0\qquad\nn{Im}w|_\Sigma=0.\label{slag}\ee
We will now specify the pure spinor equations and the equations of motion for both the cases with and without a DWSB contribution.
\subsubsection{Backgrounds with only SSB contributions}
We start by discussing the class of backgrounds having only an SSB contribution. The domain-wall BPSness \eqref{retSSB} first imposes
\be \dd(e^{3A-\phi}\theta)=0.\label{Aphi}\ee
This means that locally, one can introduce a complex coordinate $z$ such that $\dd z=e^{3A-\phi}\theta$. Then, the hypersurface $D$ defined by $z=\nn{constant}$ admits an SU$(2)$ structure defined by the pair $(\mathfrak{j}|_D,w|_D)$, and \eqref{slag} means that the fibre $\Sigma$ defines a Slag fibration of the leaves $D$.

    We can choose a local basis such that $e^1,e^2,e^4,e^5$ are along D, and $e^1,e^2$ are tangent to the fibre $\Sigma$. We also take
    \begin{align}
        \mathfrak{j}=&-(e^1\wedge e^4+e^2\wedge e^5)\\
        w=&\ (e^1+ie^4)\wedge(e^2+i e^5)\\
        \theta=&\ e^3+ie^6.
    \end{align}

    The domain-wall BPSness also imposes
    \be \dd\mathfrak{j}|_D=0\qquad\quad H|_D=0,\label{jH}\ee
   so one can expand $\dd \mathfrak{j}$ and $H$ as
   \begin{align}
       \dd\mathfrak{j}=&\ (f\wedge\bar{\theta}+c.c.)+\frac{i}{2}u\wedge\theta\wedge\bar{\theta}\label{dj}\\
       H=&\ (g\wedge\bar{\theta}+c.c.)+\frac{i}{2}h\wedge\theta\wedge\bar{\theta}\label{Hparam}
   \end{align}
   with $f$ and $g$ complex two-forms, and $u$ and $h$ real one-forms which can be decomposed purely along $D$. The last condition imposed by domain-wall BPSness is then
   \be g+if=0.\label{gf}\ee
   The gauge BPSness yields the following RR-fluxes
   \begin{align}
       e^\phi\ast F_1=&\ H\wedge\nn{Re}w-i\dd(2A-\phi)\wedge\nn{Im}w\wedge\theta\wedge\bar{\theta}\\
       \ast F_3=&-e^{-4A}\dd(e^{4A-\phi}\nn{Re}w)\\ 
       F_5=&\ 0,
   \end{align}
   and the violation of the D-string BPSness takes the form
   \begin{align}
       &\dd(e^{2A-\phi}\nn{Im}w)=0\label{su2dsbps1}\\
       &e^{2A-\phi}\big[H\wedge\nn{Im}w-\frac{i}{2}\dd(\nn{Re}w\wedge \theta\wedge\bar{\theta})\big]=4\pi\sqrt{\alpha'}e^A(R_1\hat{\alpha}_1\dd y^1+R_2\hat{\alpha}_2\dd y^2)\wedge j\label{su2dsbps2}
   \end{align}
   with
\be j=-4e^{3A-\phi}\nn{vol}_{\mathcal{B}_4}.\ee
Turning to the equations of motion, the NSNS-field equation is
\be \dd\Big[e^{5A-2\phi}\braket{\nn{Im}w,\hat{\alpha}^a\iota_a(\nn{Re}w-\nn{Re}w|_\Sigma)}\Big]=0\qquad a=1,2\label{NSSU2}\ee
while the internal Einstein equations reduce to
\be \braket{g_{k(m}\dd y^k\wedge\iota_{n)}\nn{Im}w,\dd(e^{3A-\phi}\hat{\alpha}^a\iota_a(i\nn{Im}w\wedge\theta\wedge\bar{\theta})}=0\qquad a=1,2.\ee
We now turn to the construction of a concrete background satisfying these conditions.

    To do so, we consider a factorisable warped six-torus
    \begin{align}
                \dd s^2=&\text{e}^{2A}\dd s^2_{\mathbb{R}_{1,3}}+  \dd s^2_{\mathcal{M}}\\
                    \dd s^2_{\mathcal{M}}=&\alpha'(2\pi)^2\Big\{\text{e}^{2A}\big[R_1^2(\dd y^1)^2+R_2^2(\dd y^2)^2\big]+\text{e}^{-2A}\sum_{j=3}^6R^2_j(\dd y^j)^2\Big\},
                \end{align}
    and we take the two-torus spanned by $y^1$ and $y^2$ to be the fibre $\Sigma$ over the four-torus base spanned by $y^3,y^4,y^5,y^6$. We also consider the fibre to be wrapped by $2^4$ O5-planes and $n_{D5}$ D5-branes. The SU$(2)$ structure is then
    \begin{align}
        \mathfrak{j}=&-\alpha'(2\pi)^2(R_1R_4\dd y^1\wedge\dd y^4+R_2R_5\dd y^2\wedge\dd y^5)\\
        w=&\ \alpha'(2\pi)^2(e^AR_1\dd y^1+i e^{-A}R_4\dd y^4)\wedge(e^AR_2\dd y^2+i e^{-A}R_5\dd y^5)\\
        \theta=&\ 2\pi\sqrt{\alpha'}e^{-A}(R_3\dd y^3+i R_6\dd y^6).
    \end{align}
    Let us now see what the domain-wall BPSness imposes on the background. First of all \eqref{Aphi} sets 
    \be e^{2A-\phi}=\nn{const.}\ee
Then, if we consider the following NSNS-field ansatz
\be H=(2\pi)^2\alpha'\big[N_{\nn{NS}1}\dd y^3\wedge\dd y^4\wedge\dd y^6+N_{\nn{NS}2}\dd y^3\wedge\dd y^5\wedge\dd y^6\big]\qquad N_{\nn{NS}1,2}\in\mathbb{Z},\ee
we see that \eqref{jH} and \eqref{gf} are satisfied, with $g=f=0$, and we have $u=0$ and $h=-\frac{e^{-2A}}{R_3R_6}\big[N_{\nn{NS}1}\dd y^4+N_{\nn{NS}2}\dd y^5\big]$.

    The generalised current is 
        \be j=-4\alpha'^2 (2\pi)^4e^{-A-\phi} R_3 R_4 R_5 R_6  \dd y^3\wedge \dd y^4\wedge \dd y^5\wedge \dd y^6.\ee
    The violation of the D-string BPSness \eqref{su2dsbps1} is identically satisfied, while \eqref{su2dsbps2} reduces to
    \be e^{2A-\phi}H\wedge\nn{Im}w=4\pi\sqrt{\alpha'}e^A(R_1\hat{\alpha}_1\dd y^1+R_2\hat{\alpha}_2\dd y^2)\wedge j\ee
    which is satisfied with
    \begin{align}
        \hat{\alpha}_1=&-\frac{e^{2A}N_{\nn{NS}1}}{16\pi\sqrt{\alpha'}R_3R_4R_6}\\
         \hat{\alpha}_2=&-\frac{e^{2A}N_{\nn{NS}2}}{16\pi\sqrt{\alpha'}R_3R_5R_6}.
    \end{align}
    The gauge BPSness constrains the RR-fluxes to be
    \begin{align}
        F_1=&e^{2A-\phi}\big[\frac{R_4N_{\nn{NS}2}}{R_3R_5R_6}\dd y^4-\frac{R_5N_{\nn{NS}1}}{R_3R_4R_6}\dd y^5\big]\\
        F_3=&e^{2A-\phi}\ast_{\mathcal{B}_4}\dd e^{-4A}
    \end{align}
    with $\ast_{\mathcal{B}_4}$ the four-dimensional Hodge operator on the unwarped base. 

        The RR-flux quantisation implies that we must have
        \be e^{2A-\phi}\frac{R_5N_{\nn{NS}1}}{R_3R_4R_6}=N_{\nn{R}1}\in\mathbb{Z}\qquad e^{2A-\phi}\frac{R_4N_{\nn{NS}2}}{R_3R_5R_6}=N_{\nn{R}2}\in\mathbb{Z},\ee
so the RR Bianchi identities reduce to
\be -\hat{\nabla}^2_{\mathcal{B}_4}e^{-4A}=\frac{1}{e^{2A-\phi}(2\pi)^2\alpha'\Pi_{a=3}^6R_a}\big[N_{\nn{NS}1}N_{\nn{R}1}+N_{\nn{NS}2}N_{\nn{R}2}+\sum_{i\in\nn{D}5\nn{'}s,\nn{O}5\nn{'}s}q_i\delta^4_{\mathcal{B}_4}(y_i)\big]\ee
with again $q_{\nn{D}5}=-q_{\nn{O}5}=1$. Integrating this condition on the base yields the following tadpole condition
\be N_{\nn{NS}1}N_{\nn{R}1}+N_{\nn{NS}2}N_{\nn{R}2}+n_{\nn{D}5}=16.\ee
Turning to the equations of motion, the NSNS-field equation \eqref{NSSU2} is identically satisfied, and we have
\be \dd(e^{3A-\phi}\hat{\alpha}^a\iota_a(i\nn{Im}w\wedge\theta\wedge\bar{\theta})=0\ee
so the internal Einstein equations are satisfied. 
\subsubsection{Backgrounds with both SSB and DWSB contributions}
We now consider SU$(2)$ backgrounds with space-filling D5-branes with both SSB and DWSB contributions. 
The internal manifold geometry is similar to the one of the backgrounds with only an SSB contribution. Indeed, the domain-wall BPSness violation \eqref{rssbdwsb} keeps on imposing
\be \dd(e^{3A-\phi}\theta)=0,\label{thetassbdwsb}\ee
so we can similarly define the hypersurface $D$ along $e^1,e^2,e^4,e^5$, which admits a Slag fibration with fibres $\Sigma$. We keep the same parametrisation of the SU$(2)$ structure
    \begin{align}
        \mathfrak{j}=&-(e^1\wedge e^4+e^2\wedge e^5)\\
        w=&\ (e^1+ie^4)\wedge(e^2+i e^5)\\
        \theta=&\ e^3+ie^6,
    \end{align}
and \eqref{rssbdwsb} also imposes
    \be \dd\mathfrak{j}|_D=0\qquad\quad H|_D=0.\label{djHssbdwsb}\ee
    However, using the same expansion of $\dd \mathfrak{j}$ and $H$ \eqref{dj}, \eqref{Hparam}, the domain-wall BPSness violation now imposes
    \be g+if=-2re^4\wedge e^5.\label{gfssbdwsb}\ee

        The gauge BPSness sets the RR-fluxes to be
         \begin{align}
       e^\phi\ast F_1=&\ H\wedge\nn{Re}w-i\dd(2A-\phi)\wedge\nn{Im}w\wedge\theta\wedge\bar{\theta}\\
       \ast F_3=&-e^{-4A}\dd(e^{4A-\phi}\nn{Re}w)\\ 
       F_5=&\ 0,
   \end{align}
   and the violation of the D-string BPSness takes the form
   \begin{align}
       &\dd(e^{2A-\phi}\nn{Im}w)=0\label{svsu21}\\
       &e^{2A-\phi}\big[H\wedge\nn{Im}w-\frac{i}{2}\dd(\nn{Re}w\wedge \theta\wedge\bar{\theta})\big]=4\pi\sqrt{\alpha'}e^A(R_1\hat{\alpha}_1\dd y^1+R_2\hat{\alpha}_2\dd y^2)\wedge j\label{svsu22}
   \end{align}
   with
\be j=-4e^{3A-\phi}\nn{vol}_{\mathcal{B}_4}.\ee
Turning to the equations of motion, the NSNS-field equation is
\be e^{3A-\phi}\nn{Im}\braket{\theta,\dd \big[e^{A-\phi}r^\ast\Tilde{\Theta}\big]}+2\dd\big[e^{5A-2\phi}\braket{\nn{Im}w,\hat{\alpha}^a\iota_a\Tilde{\Theta}}\big]=0\qquad a=1,2,\label{NSSU2sd}\ee
with $\Tilde{\Theta}=\nn{Re}w-\nn{Re}w|_\Sigma$. Specifying the internal Einstein equations \eqref{IEEqssbdwsb} to this case gives a rather long and not particularly enlightening expression, so we just give here the following conditions
\begin{align}
    &\dd\big[e^{3A-\phi}\hat{\alpha}^a\iota_a(i\nn{Im}w\wedge\theta\wedge\bar{\theta})\big]=0\\
    &\dd \big[e^{A-\phi}r^\ast i\nn{Im}w\wedge\theta\wedge\bar{\theta}\big]=0\\
     &\dd \big[e^{A-\phi}r^\ast\Tilde{\Theta}\big]=0\label{IEEQsu2sd}
\end{align}
which are stronger than the internal Einstein equations, but are reasonable conditions to impose on such SU$(2)$ backgrounds, and which guarantee that the internal Einstein equations are satisfied. The example background that we construct below will satisfy these conditions.

    We consider here again a factorisable six-torus
\begin{align}
                \dd s^2=&\text{e}^{2A}\dd s^2_{\mathbb{R}_{1,3}}+  \dd s^2_{\mathcal{M}}\\
                    \dd s^2_{\mathcal{M}}=&\alpha'(2\pi)^2\Big\{\text{e}^{2A}\big[R_1^2(\dd y^1)^2+R_2^2(\dd y^2)^2\big]+\text{e}^{-2A}\sum_{j=3}^6R^2_j(\dd y^j)^2\Big\},
                \end{align}
    and as before we take the two-torus spanned by $y^1$ and $y^2$ to be the fibre $\Sigma$ over the four-torus base spanned by $y^3,y^4,y^5,y^6$. We also consider the fibre to be wrapped by $n_{O5}$ O5-planes and $n_{D5}$ D5-branes. The SU$(2)$ structure is again
    \begin{align}
        \mathfrak{j}=&-\alpha'(2\pi)^2(R_1R_4\dd y^1\wedge\dd y^4+R_2R_5\dd y^2\wedge\dd y^5)\\
        w=&\ \alpha'(2\pi)^2(e^AR_1\dd y^1+i e^{-A}R_4\dd y^4)\wedge(e^AR_2\dd y^2+i e^{-A}R_5\dd y^5)\\
        \theta=&\ 2\pi\sqrt{\alpha'}e^{-A}(R_3\dd y^3+i R_6\dd y^6).
    \end{align}
From \eqref{thetassbdwsb}, we see that the domain-wall BPSness violation sets again
\be e^{2A-\phi}=\nn{const.}\ee
However, we now consider the following NSNS-field
\begin{align} H=(2\pi)^2\alpha'&\big[N_{\nn{NS}1}\dd y^3\wedge\dd y^4\wedge\dd y^6+N_{\nn{NS}2}\dd y^3\wedge\dd y^5\wedge\dd y^6\nonumber\\
&+N_{\nn{NS}3}\dd y^4\wedge\dd y^5\wedge\dd y^6\big]\qquad N_{\nn{NS}1,2,3}\in\mathbb{Z},\label{Hsu2ssbdwsb}
\end{align}
and we see that \eqref{djHssbdwsb} keep on being satisfied. We now have $f=u=0$ and $g=\frac{i\pi\sqrt{\alpha'}}{R_6}e^AN_{\nn{NS}3}\dd y^4\wedge\dd y^5$, and the domain-wall BPSness violation \eqref{gfssbdwsb} now reads
\be g+if=-4\pi r\sqrt{\alpha'}e^{-2A}R_4R_5\dd y^4\wedge\dd y^5\ee
with
\be r=-\frac{ie^{3A}N_{\nn{NS}3}}{8\pi\sqrt{\alpha'}R_4R_5R_6}.\ee
  The additional DWSB contribution, which happens through the new components of the NSNS-flux in \eqref{Hsu2ssbdwsb}, doesn't modify the violation of the D-string BPSness: \eqref{svsu21} is again identically satisfied, while \eqref{svsu22} reduces to
    \be e^{2A-\phi}H\wedge\nn{Im}w=4\pi\sqrt{\alpha'}e^A(R_1\hat{\alpha}_1\dd y^1+R_2\hat{\alpha}_2\dd y^2)\wedge j\ee
    which is satisfied with
    \begin{align}
        \hat{\alpha}_1=&-\frac{e^{2A}N_{\nn{NS}1}}{16\pi\sqrt{\alpha'}R_3R_4R_6}\\
         \hat{\alpha}_2=&-\frac{e^{2A}N_{\nn{NS}2}}{16\pi\sqrt{\alpha'}R_3R_5R_6}.
    \end{align}
The gauge BPSness sets the RR-fluxes to 
\begin{align}
        F_1=&e^{2A-\phi}\big[\frac{R_4N_{\nn{NS}2}}{R_3R_5R_6}\dd y^4-\frac{R_5N_{\nn{NS}1}}{R_3R_4R_6}\dd y^5-\frac{R_3N_{\nn{NS}3}}{R_4R_5R_6}\dd y^3\big]\\
        F_3=&e^{2A-\phi}\ast_{\mathcal{B}_4}\dd e^{-4A}
    \end{align}
 The RR-flux quantisation implies that we must have
        \begin{align}
        e^{2A-\phi}\frac{R_5N_{\nn{NS}1}}{R_3R_4R_6}=N_{\nn{R}1}\qquad e^{2A-\phi}\frac{R_4N_{\nn{NS}2}}{R_3R_5R_6}=N_{\nn{R}2}\qquad  e^{2A-\phi}\frac{R_3N_{\nn{NS}3}}{R_4R_5R_6}=N_{\nn{R}3}
        \end{align}
        with $N_{\nn{R}1},N_{\nn{R}2},N_{\nn{R}3}\in\mathbb{Z}$.
The RR Bianchi identities then reduce to
\be -\hat{\nabla}^2_{\mathcal{B}_4}e^{-4A}=\frac{1}{e^{2A-\phi}(2\pi)^2\alpha'\Pi_{a=3}^6R_a}\big[\sum_{k=1}^3N_{\nn{NS}k}N_{\nn{R}k}+\sum_{i\in\nn{D}5\nn{'}s,\nn{O}5\nn{'}s}q_i\delta^4_{\mathcal{B}_4}(y_i)\big]\ee
with again $q_{\nn{D}5}=-q_{\nn{O}5}=1$. Integrating this condition on the base yields the following tadpole condition
\be \sum_{k=1}^3N_{\nn{NS}k}N_{\nn{R}k}+n_{\nn{D}5}=16.\ee
Turning to the equations of motion, the conditions \eqref{NSSU2sd}$-$\eqref{IEEQsu2sd}  are obeyed, provided that $e^{2A-\phi}=\nn{constant}$. The NSNS-field equation and internal Einstein equations are therefore satisfied, as discussed above.

\section{Conclusion}

 In this paper, we investigated some corners of the landscape of non-supersymmetric flux vacua, in the light of Generalised Complex Geometry. 
In Generalised Complex Geometry the conditions for warped backgrounds with $\mathcal{N}=1$ supersymmetry can be expressed as differential equations on polyforms, called pure spinors, which, in turn, are
seen as generalised calibration conditions for various types of D-branes wrapping cycles on the internal manifold: D-string, domain-wall, or space-filling branes, from the four-dimensional perspective. 

This interpretation provides different controlled ways of breaking supersymmetry in ten-dimensions by violating the BPSness of D-string, domain-wall, or space-filling probe branes. 
The philosophy is to look at non-supersymmetric
solutions via a two-step procedure: one first solves the deformed supersymmetry equations, which are still first order equations, and then look at the extra constraints that the equations of motion impose. 

This approach was first used in  \cite{Lust:2008zd}, where non-supersymmetric solutions were found by deforming the domain-wall calibration condition. 
Our work extends \cite{Lust:2008zd} to new classes of solutions obtained by deforming the D-string BPS condition or both the domain-wall and D-string ones. 

To solve the ten-dimensional equations of motion, it proves convenient to use an effective four-dimensional potential obtained by integrating the ten-dimensional supergravity action on the internal manifold. The effective potential can be expressed in terms of  pure spinors as an integral over the compact space. Then the equations of motion are obtained by varying it  with respect to the physical fields and 
are also given in terms of the pure spinors. The equations of motion are rather convoluted in the general case and are still to be studied. However, by choosing specific classes of supersymmetry breaking, they become tractable.  
We were able to solve them for a variety of concrete type II SU$(2)$ and SU$(3)$ backgrounds respecting \eqref{SSBcoord}.  
        
When studying the effective potential for the SSB backgrounds, we witnessed the presence of some terms belonging to vector representations of the SU$(3)\times $SU$(3)$ structure, which are believed to be massive modes from the effective point of view. In that sense, our classes of backgrounds describe a set of fully ten-dimensional non-supersymmetric solutions of type II supergravity. It is not clear weather they admit a proper four-dimensional interpretation.  It would be interesting to apply the recently developed exceptional generalised geometry techniques \cite{Josse} to study eventual consistent truncations of such backgrounds.

However, we showed that our class of background shares a property with the one-parameter DWSB class introduced in \cite{Lust:2008zd}, namely the existence of a `truncation' dictated by the geometry such that the `off-shell' effective potential is positive semi-definite, and vanishes at the solutions. By `truncation' we meant that we only consider the off-shell deformations of the potential which are compatible with the generalised foliation of the internal space, so it is by no mean a rigorous truncation to a finite set of mode. We therefore present this as an interesting property of our class of backgrounds, but this is a weaker statement than arguing for the perturbative stability of our backgrounds, since we have no control over the modes we are `truncating' away and keeping in our `off-shell' potential.
            
            We carried the same analysis, from the effective potential and the derivation of the equations of motion, to the construction of concrete backgrounds, for another class of vacua, which have the same supersymmetry breaking term violating the D-string BPSness, but supplemented with the one-parameter DWSB contribution violating the domain-wall BPSness. Each supersymmetry breaking term brings different contributions to the effective potential and therefore to the equations of motion, which can be solved separately, which we did for several explicit SU$(2)$ and SU$(3)$ background constructions.

\medskip
            
One obvious extension of this work would be to consider different patterns of the D-string BPSness violation, depending on the generalised current associated to the background D-branes or not, and to look for explicit solutions. 

Another natural extension of the present work is to carry the constructions of the analogous Heterotic backgrounds, where the building block of the supersymmetry breaking term entering the modified D-string condition could now be, for instance, the base volume-form of some elliptically fibered internal manifolds like the ones discussed in \cite{DWSBheterotic}.

Another interesting direction would be to consider solutions of ten-dimensional supergravity violating the D-string BPSness, but without any vector modes under the SU$(3)\times $SU$(3)$ structure, for which it would probably be possible to get a deeper understanding of their associated effective theories. In the case of vacua with an external space being Minkowski, such a D-string BPSness violation should be accompanied by a domain-wall BPSness violation, otherwise there would only be positive contributions to the effective potential. One could then hope to interpret the associated effective theories as, for instance, solutions of four-dimensional $\mathcal{N}=1$ supergravity with non-vanishing D-terms, F-terms and superpotential, extending the supersymmetric analysis of \cite{Martucci}. We will come back to these questions in \cite{Menet:2023rml}.

    Finally, the techniques from Exceptional Generalised Geometry can shed a different light on the study of non-supersymmetric equations of motion: by classifying the possible supersymmetry breaking terms in terms of different representations of the torsion of some Generalised structure, one can reformulate the equations of motion as first order differential conditions on these torsion representations. We hope to come back to this idea in the near future \cite{Menet:2024}.

\begin{center}
    \textbf{Acknowledgements}
\end{center}
It is a pleasure to thank Davide Cassani, Mariana Gra{\~n}a, Luca Martucci, Michela Petrini and Dan Waldram for insightful discussions.

  \appendix
\section{Supergravity Conventions}\label{sec:AppA}
\subsection{Bosonic sector}

Our supergravity conventions are identical to the ones of \cite{Lust:2008zd}. We introduce the following ten- and six-dimensional Hodge operators
\begin{align}
\Tilde{\ast}_{10}=&\ast_{10}\circ\sigma\\
\Tilde{\ast}_6=&\ast_{6}\circ\sigma
\end{align} 
with, for a $p$-form $\omega$
\begin{align}
&\ast_{10}\omega_p=-\frac{1}{p!(10-p)!}\sqrt{-g}\ \epsilon_{M_1...M_{10}}\omega^{M_{11-p}...M_{10}}\dd x^{M_1}\wedge ...\wedge \dd x^{M_{10-p}}\\
&\ast_{6}\omega_p=\frac{1}{p!(6-p)!}\sqrt{-g}\ \epsilon_{m_1...m_{6}}\omega^{m_{7-p}...m_{6}}\dd y^{m_1}\wedge ...\wedge \dd 
y^{m_{6-p}}.\end{align}
The ten-dimensional RR-field strength self duality is then
\be F^{10}=\Tilde{\ast}_{10}F^{10}.\ee
The type II pseudo-action in democratic formalism is
\be S=\frac{1}{2\kappa^2_{10}}\int \dd^{10}x\sqrt{-g}\Big\{e^{-2\phi}[R+4(\dd\phi)^2-\frac{1}{2}H^2]-\frac{1}{4}F^2\Big\}+S^{(\nn{loc})},\ee
where $2\kappa^2_{10}=(2\pi)^7\alpha'^4$ and for any real p-form $\omega$ we define $\omega^2=\omega\cdot\omega$ with $\cdot$ defined as
\be \omega\cdot\chi=\frac{1}{p!}\omega_{M_1...M_p}\chi^{M_1...M_p}.\ee
In the text, if $\omega$ is complex, then we consider $|\omega|^2=\omega\cdot\bar{\omega}$.
Varying this action and imposing the self-duality conditions, we find the following equations of motion.

    The dilaton equation 
    \be \nabla^2\phi-(\dd\phi)^2+\frac{1}{4}R-\frac{1}{8}H^2-\frac{1}{4}\frac{\kappa^2_{10}e^{2\phi}}{\sqrt{-g}}\frac{\delta S^{(\nn{loc})}}{\delta\phi}=0,\ee
the $B$-field equation
    \be -\dd(e^{-2\phi}\ast_{10}H)+\frac{1}{2}[\ast_{10}F\wedge F]_8+2\kappa^2_{10}\frac{\delta S^{(\nn{loc})}}{\delta B}=0,\ee
the Einstein equation 
\begin{align}
    &e^{-2\phi}[g_{MN}+2g_{MN}\dd\phi\cdot\dd\phi-2g_{MN}\nabla^2\phi+2\nabla_M\nabla_N\phi\nonumber\\
    &-
    \frac{1}{2}\iota_MH\cdot\iota_NH+\frac{1}{4}g_{MN}H\cdot H]-\frac{1}{4}\iota_MF\cdot\iota_NF-\kappa^2_{10}T^{(\nn{loc})}_{MN}=0,
\end{align}
with \be T^{(\nn{loc})}_{MN}=-\frac{2}{\sqrt{-g}}\frac{\delta S^{(\nn{loc})}}{g^{MN}},\ee   and the RR-fluxes variation gives the Bianchi identities
\be \dd_HF=-j_\nn{source}.\ee
Combining the dilaton equation of motion with the Einstein equations, one can write the modified Einstein equations
\begin{align}
    &R_{MN}+2\nabla_M\nabla_N\phi-\iota_MH\cdot\iota_NH-\frac{1}{4}e^{2\phi}\iota_MF\cdot\iota_NF\\
    &-\kappa^2_{10}e^{2\phi}\Big(T^{(\nn{loc})}_{MN}+\frac{g_{MN}}{2\sqrt{-g}}\frac{\delta S^{(\nn{loc})}}{\delta\phi}\Big)=0\label{modeinstein}.
\end{align}
Finally, we define the Mukai pairing for a pair of polyforms $\omega$ and $\chi$
\be \braket{\omega,\chi}=\omega\wedge\sigma(\chi)|_6,\ee
and more generally, we use throughout the paper
\be \braket{\omega,\chi}_k=\omega\wedge\sigma(\chi)|_k.\ee
In the case of a six-dimensional manifold $\mathcal{M}$, the Mukai pairing satisfies the following property
\be \int_\mathcal{M}\braket{\dd_H\omega,\chi}=\int_\mathcal{M}\braket{\omega,\dd_H\chi}\label{Mukaid}.\ee
\subsection{Gamma matrices}
We use a real representation of the ten-dimensional gamma matrices $\Gamma_M$. The ten-dimensional chiral operator is
\be \Gamma_{(10)}=\Gamma^{01...9}\ee
with flat ten-dimensional indices. For any p-form $\omega$, we denote its image under the Clifford map $\slashed{\omega}$ with
\be \omega\equiv\frac{1}{p!}\omega_{M_1...M_p}\dd x^{M_1...M_p}\quad \longleftrightarrow\quad\slashed{\omega}=\frac{1}{p!}\omega_{M_1...M_p}\Gamma^{M_1...M_p}.\label{cliffmap}\ee

    We define the splitting of the ten-dimensional gamma matrices into four- and six-dimensional gamma matrices $\hat{\gamma}^\mu$ and $\gamma^m$ as
    \be \Gamma^\mu=e^{-A}\hat{\gamma}^\mu\otimes\mathbb{1}\qquad\Gamma^m=\gamma_{(4)}\otimes\gamma^m.\ee
The $\hat{\gamma}^\mu$ are associated to the unwarped four-dimensional metric, and $\gamma_{(4)}=i\hat{\gamma}^{0123}$ is the usual four-dimensional chiral operator. The six-dimensional chiral operator is $\gamma_{(6)}=-i\gamma^{123456}$ so we have $\Gamma_{(10)}=\gamma_{(4)}\otimes\gamma_{(6)}$. 

    The chirality of the internal spinors is then
    \be \gamma_{(6)}\eta_1=\eta_1\qquad\gamma_{(6)}\eta_2=\mp\eta_2\qquad\nn{in type IIA/IIB}.\ee

\section{Some Generalised Complex Geometry elements}
\label{sec:AppB}

                    \hspace{0.5cm}Generalised vectors are sections of the generalised tangent bundle, locally defined as $E\simeq T\mathcal{M}\oplus T^\ast\mathcal{M}$. The generalised tangent bundle admits a natural SO$(6,6)$ structure, defined by the inner product on generalised vectors
   \be V\cdot W=\iota_v\rho+\iota_w\xi,\ee
    with $V=v+\xi\in\Gamma(E)$ and $W=  w+\rho\in\Gamma(E)$. Using a two-component notation for the generalised
vectors
\be V=v+\xi\equiv \begin{pmatrix}
v\\
\xi
\end{pmatrix},\ee
one can parametrise the SO$(6,6)$ generators as
\be \mathcal{O}_A=\begin{pmatrix}
\ A & 0\\
\ 0 & (A^T)^{-1}
\end{pmatrix}\qquad\mathcal{O}_b=\begin{pmatrix}
\ \mathbb{1} &\ \ 0\ \\
\ b &\ \ \mathbb{1}\ 
\end{pmatrix}\qquad\mathcal{O}_\beta=\begin{pmatrix}
\ \mathbb{1} &\ \ \beta\ \\
\ 0 &\ \ \mathbb{1}\ 
\end{pmatrix},\ee
with $A$ any GL$(d,\mathbb{R})$ matrix, $b$ any two-form and $\beta$ any two-vector.
The SO$(6,6)$ adjoint action on generalised vectors is then
\begin{align}
    V=v+\xi\quad\rightarrow\quad\mathcal{O}_A\cdot V=&\ Av+(A^T)^{-1}\xi\\
    \quad\mathcal{O}_b\cdot V=&\ v+(\xi-\iota_v b)\\
     \quad\mathcal{O}_\beta\cdot V=&\ (v-\iota_\beta \xi)+\xi.
\end{align}
    
    On another note, the generalised vectors obey the Cliff$(6,6)$ Clifford algebra
\be \{V,W\}=(V, W)\qquad V,\ W\in\Gamma(E),\ee
and the Spin$(6,6)$ spinors are sections of a spinor representation of Cliff$(6,6)$, locally isomorphic to the space of polyforms. The action of a generalised vector $V=v+\xi$ on such a spinor $\Psi$ is
\be V\cdot \Psi=\iota_v\Psi+\xi\wedge\Psi.\label{vecspinoract}\ee
A line-bundle of pure spinors is in one-to-one correspondence with an (almost) generalised complex structure, see for instance \cite{Gualtieri:2007ng} for more details.

    An almost generalised complex structure is a map
    \be \mathcal{J}\ :\ E\rightarrow E,\ee
respecting
\begin{align}
    \mathcal{J}^2=&-\mathbb{1}\\
    \mathcal{J}^T\mathcal{I}\mathcal{J}=&\mathcal{I}
\end{align}
with
\be \mathcal{I}\ : E\times E\rightarrow \mathbb{R}\quad\nn{such that}\quad\mathcal{I}(V,W)=(V, W)\qquad V,\ W\in\Gamma(E).\ee
It is then an integrable almost generalised complex structure or simply generalised complex structure if its $+i-$eigenbundle is stable under the following Courant bracket
\be [V,W]_C=[v,w]+\mathcal{L}_v\rho-\mathcal{L}_w\xi-\frac{1}{2}\dd(\iota_v\rho-\iota_w\xi),\ee
with $V=v+\xi\in\Gamma(E)$ and $W=  w+\rho\in\Gamma(E)$. An almost generalised complex structure is $H$-integrable if its $+i-$eigenbundle is stable under the twisted Courant bracket
\be [V,W]^H_C=[V,W]_C+\iota_v\iota_wH.\label{twistcour}\ee
When considering the Clifford action on a polyform $\omega$, one can write the twisted Courant bracket as 
\be [V,W]^H_C\cdot\omega=[\{V\cdot,\dd_H\},W\cdot]\omega.\ee
One can also define a natural action of (almost) generalised complex structures on the space of differential forms, explicitly\footnote{See \cite{Cavalcanti} for the formal details.}:
\be \mathcal{J}\cdot=\frac{1}{2}\big(J_{mn}\dd y^m\wedge \dd y^n\wedge+2 {I^m}_n[\dd y^n,\iota_m]+P^{mn}\iota_m\iota_n\big)\ee
with
\be \mathcal{J}=\begin{pmatrix}
    I & P\\
    J & -I^T
\end{pmatrix}.\ee
One can then decompose the space of polyforms in terms of $\mathcal{J}$ eigenbundles
\be \Lambda^\bullet T^\ast\mathcal{M}\otimes\mathbb{C}=\bigoplus_{k=-3}^3 V_k,\ee
with $k$ the eigenvalues of the sub-spaces $V_k$, which are representations of the $SU(3,3)$ structure group associated with $\mathcal{J}$. 

    For example, in the case of the (almost) generalised complex structure associated to the pure spinor $\Psi\propto\Omega$, its relation with the standard Hodge decomposition is
\be V_k=\bigoplus_p \Lambda^{3-p,3-k-p},\ee
where $\Lambda^{3-p,3-k-p}$ are sections of the $(p,q)$-forms defined by the standard associated (almost) complex structure. 

    If the generalised manifold admits two compatible pure spinors, or equivalently two commuting (almost) generalised complex structures $\mathcal{J}_1$ and $\mathcal{J}_2$, one can further decompose the space of polyforms in terms of SU$(3)\times $SU$(3)$ representations which are the common $\mathcal{J}_1$ and $\mathcal{J}_2$ eigenbundles
\be \Lambda^\bullet T^\ast\mathcal{M}\otimes\mathbb{C}=\bigoplus_{k=-3}^3\bigoplus_{l=-3}^3 V_{k,l}.\ee
This decomposition is the generalised Hodge diamond decomposition, see for instance \cite{Martucci} for more details.
\section{Supersymmetry breaking and pure spinors}
\label{sec:AppC}
We consider a ten-dimensional SU$(3)\times $SU$(3)$ background with a Minkowski four-dimensional space and a ten-dimensional bispinor $\epsilon=(\epsilon_1,\epsilon_2)^T$ as in \eqref{10dspinors}. We give here the parametrisation of the most general supersymmetry breaking, and we start by writing down the non-vanishing supersymmetry variations
 \begin{align}
      \delta\psi^{(1)}_\mu=\frac{1}{2}e^A \hat{\gamma}_\mu \zeta\otimes \mathcal{V}_1+c.c.\qquad&\delta\psi^{(2)}_\mu=\frac{1}{2}e^A \hat{\gamma}_\mu \zeta\otimes \mathcal{V}_2+c.c.\\
      \delta\psi^{(1)}_m=\zeta\otimes \mathcal{U}^1_m+c.c.\qquad&\delta\psi^{(2)}_m=\zeta\otimes \mathcal{U}^2_m+c.c.\\
      \Delta\epsilon_1=\zeta\otimes\mathcal{S}_1+c.c.\qquad&\Delta\epsilon_2=\zeta\otimes\mathcal{S}_2+c.c.,
  \end{align}
where $\mathcal{V}_{1,2},\ \mathcal{U}^{1,2}_m$ and $\mathcal{S}_{1,2}$ are internal spinors parametrising the supersymmetry breaking
\begin{align}
\mathcal{V}_1=&\ \slashed{\partial}A\eta_1+\frac{1}{4}e^{\phi}\gamma_{(6)}\slashed{F}\eta_2\nonumber\\
\mathcal{V}_2=&\ \slashed{\partial}A\eta_2-\frac{1}{4}e^{\phi}\gamma_{(6)}\slashed{F}^\dagger\eta_1\nonumber\\
\mathcal{S}_1=&\ (\slashed{\nabla}-\slashed{\partial}\phi+2\slashed{\partial}A+\frac{1}{4}\slashed{H})\eta_1\nonumber\\
\mathcal{S}_2=&\ (\slashed{\nabla}-\slashed{\partial}\phi+2\slashed{\partial}A+\frac{1}{4}\slashed{H})\eta_2\nonumber\\
\mathcal{U}^1_m=&\ (\nabla_m+\frac{1}{4}\iota_m\slashed{H})\eta_1+\frac{1}{8}e^\phi\slashed{F}\gamma_m\gamma_{(6)}\eta_2\nonumber\\
\mathcal{U}^2_m=&\ (\nabla_m-\frac{1}{4}\iota_m\slashed{H})\eta_2-\frac{1}{8}e^\phi\slashed{F}\gamma_m\gamma_{(6)}\eta_1\label{moddilatino}.
\end{align}
Following \cite{Lust:2008zd}, we expand them in terms of supersymmetry breaking parameters in the following way
\begin{align}
    \mathcal{V}_1=r_1\eta_1^\ast+s^1_m\gamma^m\eta_1&\qquad \mathcal{V}_2=r_2\eta_2^\ast+s^2_m\gamma^m\eta_2\\
\mathcal{S}_1=t_1\eta_1^\ast+u^1_m\gamma^m\eta_1&\qquad\mathcal{S}_2=t_2\eta_2^\ast+u^2_m\gamma^m\eta_2\\
     \mathcal{U}^1_m=p^1_m \eta_1+q^1_{mn}\gamma^n\eta_1^\ast&\qquad\mathcal{U}^2_m=p^2_m \eta_2+q^2_{mn}\gamma^n\eta_2^\ast.
\end{align}
It has been shown in \cite{Lust:2008zd} that these parameters do not mix under T-duality. One can now rewrite the most general non-supersymmetric pure spinor equations, expanded on the generalised Hodge diamond, in terms of these supersymmetry breaking parameters
\begin{align}
    &e^{-2A+\phi}\dd_H(e^{2A-\phi}\Psi_1)+2\dd A\wedge\nn{Re}\Psi_1-e^\phi\Tilde{\ast} _6F=\Upsilon\\
    &e^{-3A+\phi}\dd_H(e^{3A-\phi}\Psi_2)=K
\end{align}
with
\begin{align}
    \Upsilon=&\frac{1}{2}(-1)^{|\Psi_1|}(r_1^\ast+t_2^\ast)\Psi_2+\frac{1}{2}(-1)^{|\Psi_1|}(r_2+t_1)\bar{\Psi}_2+\frac{1}{2}(s^1_m)^\ast\gamma^m\bar{\Psi}_1+\frac{1}{2}(-1)^{|\Psi_1|}s^2_m\bar{\Psi}_1\gamma^m\nonumber\\
    &+\frac{1}{2}[u^1_m+(p^2_m)^\ast]\gamma^m\Psi_1+\frac{1}{2}(-1)^{|\Psi_1|}[(u^2_m)^\ast+p^1_m]\Psi_1\gamma^m\nonumber\\
    &+\frac{1}{2}(q^2_{mn})^\ast\gamma^m\Psi_2\gamma^n-\frac{1}{2}q^1_{mn}\gamma^n\bar{\Psi}_2\gamma^m\nonumber\\
    K=&\frac{1}{2}(-1)^{|\Psi_1|}t_2\Psi_2-\frac{1}{2}(-1)^{|\Psi_1|}t_1\bar{\Psi}_2+\frac{1}{2}(u^1_m+p^2_m)\gamma^m\Psi_2+\frac{1}{2}(-1)^{|\Psi_2|}(u^2_m+p_m^1)\Psi_2\gamma^m\nonumber\\
    &+\frac{1}{2}q^1_{mn}\gamma^n\bar{\Psi}_1\gamma^m-\frac{1}{2}q^2_{mn}\gamma^m\Psi_1\gamma^n.
\end{align}
Let us now specify this expansion to the most general case of pure string-like supersymmetry breaking. We first impose domain-wall BPSness, namely
\be K=0,\ee
which gives the following constraints on the supersymmetry breaking parameters
\begin{align}
    &t_1=t_2=0\\
    &u^1_m=-\frac{1}{2}{(1+iJ_1)^k}_mp^2_k\\
     &u^2_m=-\frac{1}{2}{(1+iJ_2)^k}_mp^1_k\\
     &{(1+iJ_2)^k}_mq^1_{kn}=0\\
    &{(1+iJ_1)^k}_mq^2_{kn}=0,
\end{align}
then we impose the gauge BPSness, which amounts to requiring $\Upsilon$ to be purely imaginary, which yields
\begin{align}
    &r_1=-r_2\equiv r\\
    &q^1_{mn}=q^2_{nm}\equiv q_{mn}\\
    &s^1_m=\frac{1}{2}{(1+iJ_1)^k}_mp^2_k-(p^2_m)^\ast\\
    &s^2_m=\frac{1}{2}{(1+iJ_2)^k}_mp^1_k-(p^1_m)^\ast.
\end{align}

Imposing these constraints, the internal spinors describing the pure D-string supersymmetry breaking read
\begin{align}
    \mathcal{V}_1=r\eta_1^\ast+(p^2_m-(p^2_m)^\ast)\gamma^m\eta_1&\qquad \mathcal{V}_2=-r\eta_2^\ast+(p^1_m-(p^1_m)^\ast)\gamma^m\eta_2\\
\mathcal{S}_1=-p^2_m\gamma^m\eta_1&\qquad\mathcal{S}_2=-p^1_m\gamma^m\eta_2\\
     \mathcal{U}^1_m=p^1_m \eta_1+q_{mn}\gamma^n\eta_1^\ast&\qquad\mathcal{U}^2_m=p^2_m \eta_2+q_{nm}\gamma^n\eta_2^\ast.
\end{align}
The most general D-string supersymmetry breaking term is therefore
\begin{align}
    \Upsilon=&\frac{1}{2}(-1)^{|\Psi_1|}(r^\ast\Psi_2-r\bar{\Psi}_2)+(\frac{1}{4}{(1-iJ_1)^k}_m(p^2_k)^\ast-\frac{1}{2}p^2_m)\gamma^m\bar{\Psi}_1\nonumber\\
    &+(-1)^{|\Psi_1|}(\frac{1}{4}{(1+iJ_2)^k}_mp^1_k-\frac{1}{2}(p^1_m)^\ast)\bar{\Psi}_1\gamma^m\nonumber\\
    &-(\frac{1}{4}{(1+iJ_1)^k}_mp^2_k-\frac{1}{2}(p^2_m)^\ast)\gamma^m\Psi_1\nonumber\\
    &-(-1)^{|\Psi_1|}(\frac{1}{4}{(1-iJ_2)^k}_mp^1_k-\frac{1}{2}p^1_m)\Psi_1\gamma^m\nonumber\\
    &+\frac{1}{2}(q_{mn})^\ast\gamma^m\Psi_2\gamma^n-\frac{1}{2}q_{mn}\gamma^m\bar{\Psi}_2\gamma^n.\label{genstring}
\end{align}
The non-supersymmetric backgrounds presented in \ref{subsec:SSB} have the following supersymmetry breaking parameters
\begin{align}
    r=&\ 0\\
    p^1_m=&\ e^A(-1)^{|\Psi_1|+1}\left[\frac{1}{2}{J_2^k}_m(-\delta^n_k+\frac{1}{2}{\Lambda^n}_k)\alpha_n+\frac{1}{4}\Lambda_{nm}\alpha_qJ_1^{qn}\right.\\
    &\left.+\frac{3i}{2}(-\delta^n_m+\frac{1}{2}{\Lambda^n}_m)\alpha_n+\frac{i}{4}{J_2^k}_m\Lambda_{nk}\alpha_qJ_1^{qn}\right]\\
    p^2_m=&\ e^A\left[(-1)^{|\Psi_1|+1}{J_1^k}_m\Lambda_{kn}\alpha_qJ_2^{nq}-\frac{1}{2}((-1)^{|\Psi_2|}\delta_m^n+(-1)^{|\Psi_1|}\frac{1}{2}{\Lambda_m}^n)\alpha_n\right.\\
    &\left. -3i(-1)^{|\Psi_1|}\Lambda_{mn}\alpha_qJ_2^{nq}-\frac{i}{2}{J_1^k}_m((-1)^{|\Psi_2|}\delta_k^n+(-1)^{|\Psi_1|}\frac{1}{2}{\Lambda_k}^n)\alpha_n\right]\\
    q_{mn}=&-\frac{e^A}{2}\alpha_p\left[(-1)^{|\Psi_1|}\Lambda_{mq}({\Omega_2^{qp}}_{n})^\ast-\Lambda_{qn}{\Omega_1^{pq}}_{m}\right],
\end{align}
   where $J_{1,2}$ are the (almost) complex structures defined by $\eta_1$ and $\eta_2$
\be {{J_{1,2}}^m}_n=\frac{i}{\lVert \eta_{1,2}\rVert}\eta_{1,2}^\dagger{\gamma^m}_n\eta_{1,2},\ee

and $\Omega_{1,2mnp}$ are the $(3,0)-$forms with respect to the (almost) complex structures defined by $\eta_1$ and $\eta_2$:

    \be \Omega_{1,2mnp}=\frac{1}{\lVert \eta_{1,2}\rVert}\eta_{1,2}^T\gamma_{mnp}\eta_{1,2}.\ee

In order to write down the supersymmetry breaking parameters of the non-supersymmetric backgrounds with both SSB and DWSB contributions presented in \ref{subsec:ssbdwsb}, one just have to add the pure DWSB parameters given in Appendix B of \cite{Lust:2008zd} to the one above.

\medskip

    The above supersymmetry breaking parameters correspond to the following decomposition of the supersymmetry breaking term in \eqref{SSBcoord} on the SU$(3)\times $SU$(3)$ structure defined by the two pure spinors
\begin{align}
         \dd_H(\text{e}^{2A-\phi}\text{Im}\Psi_1)=&\text{e}^{3A-\phi}\left(\gamma^m\Psi_1\left[(-1)^{|j|+1}\Lambda_{mn}\alpha_pJ_2^{np}-\frac{i}{2}\alpha_n((-1)^{|\Psi_2|}\delta^n_m+(-1)^{|j|}\frac{1}{2}{\Lambda_m}^{n})\right]\right. \nonumber\\
&+\gamma^m\bar{\Psi}_1\left[(-1)^{|j|+1}\Lambda_{mn}\alpha_pJ_2^{np}+\frac{i}{2}\alpha_n((-1)^{|\Psi_2|}\delta^n_m+(-1)^{|j|}\frac{1}{2}{\Lambda_m}^{n})\right]\nonumber\\
&+\Psi_1\gamma^m\left[-\frac{1}{4}\Lambda_{nm}\alpha_pJ_1^{pn}-\frac{i}{2}\alpha_n(-\delta^n_m+\frac{1}{2}{\Lambda^n}_{m})\right]\nonumber\\
&+\bar{\Psi}_1\gamma^m\left[-\frac{1}{4}\Lambda_{nm}\alpha_pJ_1^{pn}+\frac{i}{2}\alpha_n(-\delta^n_m+\frac{1}{2}{\Lambda^n}_{m})\right]\nonumber\\
&+\frac{i}{4}\gamma^m\Psi_2\gamma^n\alpha_p\left[(-1)^{|j|}\Lambda_{mq}({\Omega_2^{qp}}_{n})^\ast-\Lambda_{qn}{\Omega_1^{pq}}_{m}\right]\nonumber\\
&\left.-\frac{i}{4}\gamma^m\bar{\Psi}_2\gamma^n\alpha_p\left[(-1)^{|j|}\Lambda_{mq}{\Omega_2^{qp}}_{n}-\Lambda_{qn}({\Omega_1^{pq}}_{m})^\ast\right]\right)\label{ssbdiamond}.
    \end{align}
    
 \bibliographystyle{JHEP}
 \bibliography{biblio}

\end{document}